\documentclass[journal]{IEEEtran}
\usepackage{cite}
\usepackage{amsmath,amssymb,amsfonts}
\usepackage{algorithmic}
\usepackage{graphicx,color}
\usepackage{latexsym}
\usepackage{multirow}
\usepackage{hyperref}
\usepackage{nomencl}
\makenomenclature
\usepackage{etoolbox}
\renewcommand\nomgroup[1]{%
  \item[\bfseries
  \ifstrequal{#1}{A}{Acronym}{%
  \ifstrequal{#1}{B}{Symbols}{}}%
]}
\usepackage[linesnumbered,ruled]{algorithm2e}



\usepackage{bbding}
\usepackage{array}
\usepackage{fixltx2e}
\usepackage{stfloats}
\usepackage{caption}
\usepackage{subfigure}
\usepackage{xcolor}
\usepackage{textcomp}
\graphicspath{{Figures/}}   % Location of your graphics files
\begin{document}

\title{OTFS-IM-Assisted Non-Terrestrial Networks Relying on Autoencoder-Aided Soft-Decision Detection}

\author{Xinyu~Feng,~\IEEEmembership{Member,~IEEE,}
        Chao~Zhang,~\IEEEmembership{Member,~IEEE,}
        Mohammed EL-Hajjar,~\IEEEmembership{Senior Member,~IEEE,} \\
        Chao Xu,~\IEEEmembership{Senior Member,~IEEE,}
        and~Lajos~Hanzo,~\IEEEmembership{Life Fellow,~IEEE}% <-this % stops a space
\thanks{The authors are with the School of Electronics and Computer Science, University of Southampton, Southampton SO17 1BJ, U.K. (e-mail: {xf1e24,cz1r24,meh,cx1g08,lh}@ecs.soton.ac.uk}.}

\maketitle
\begin{abstract}
Orthogonal Time Frequency Space ({OTFS}) modulation offers significant advantages over Orthogonal Frequency Division Multiplexing ({OFDM}), particularly in high speed environments.  Hence, we consider {OTFS} transmission over high-Doppler Non-Terrestrial Networks ({NTN}). However, OTFS-based systems inherit some deficiencies from {OFDM}, such as its high peak to average power ratio, the bandwidth efficiency loss due to the cyclic prefix, and the sensitivity to the carrier frequency offset. Against this background, we harness both Multi-Band Discrete Fourier Transform-based Spreading (MB-DFT-S) and Index Modulation ({IM}) in our {OTFS} system, termed as MB-DFT-S-OTFS-IM. More explicitly, 1) DFT-S has been shown to reduce the {PAPR}; 2) {IM} is capable of improving the throughput by harnessing it in the Delay and Doppler ({DD}) domain; and 3) MB-DFT-S-OTFS-IM provides frequency diversity gain, which  benefits the tolerance to  carrier frequency offset.  Furthermore, we propose a {PAPR} reduction method based on a Deep Learning ({DL})  Autoencoder ({AE}) architecture for both hard- and soft-decision detection, where the encoder is specifically trained for minimizing {PAPR} and the decoder is conceived for accurately reconstructing the transmitted signal. Finally, we extend the proposed {AE}-aided {OTFS-IM} scheme constructed for a practical {NTN} channel model, representing a variety of satellite-to-ground schemes.

\end{abstract}


\maketitle

\nomenclature[A]{4G,5G,6G}{$4^{th}$ Generation, $5^{th}$ Generation, $6^{th}$ Generation}
\nomenclature[A]{AE}{Autoencoder}
\nomenclature[A]{AEE‑JDDIM‑OTFS}{Autoencoder‑Enhanced JDDIM‑OTFS}
\nomenclature[A]{AWGN}{Additive white Gaussian Noise}
\nomenclature[A]{BER}{Bit Error Ratio}
\nomenclature[A]{BN}{Batch Normalization}
\nomenclature[A]{CCDF}{Complementary Cumulative Distribution Function}
\nomenclature[A]{CDF}{Cumulative Distribution Function}
\nomenclature[A]{CE}{Channel Estimation}
\nomenclature[A]{CFO}{Carrier Frequency Offsets}
\nomenclature[A]{CSI}{Channel State Information}
\nomenclature[A]{DD}{Delay-Doppler}
\nomenclature[A]{DFT}{Discrete Fourier Transform}
\nomenclature[A]{DFT-S-OFDM}{Discrete Fourier Transform Spread-{OFDM}}
\nomenclature[A]{DL}{Deep Learning}
\nomenclature[A]{DNN}{Deep Neural Network}
\nomenclature[A]{FC}{Fully Connect}
\nomenclature[A]{HD}{Hard-Decision}
\nomenclature[A]{IDFT}{Inverse Discrete Fourier Transform}
\nomenclature[A]{ISFFT}{Inverse Symplectic Finite Fourier Transform}
\nomenclature[A]{IM}{Index Modulation}
\nomenclature[A]{JDDIM‑OTFS}{Joint {DD}-{IM} with OTFS}
\nomenclature[A]{LDPC}{Low-Density Parity-Check}
\nomenclature[A]{LEO}{Low Earth Orbit}
\nomenclature[A]{LLR}{Log Likelihood Ratio}
\nomenclature[A]{LTE}{Long-Term-Evolution}
\nomenclature[A]{LoS}{Line-of-Sight}
\nomenclature[A]{MAP}{Maximum \textit{a posteriori}}
\nomenclature[A]{MB}{Multi-Band}
\nomenclature[A]{MIMO}{Multiple-Input-Multiple-Output}
\nomenclature[A]{MMSE}{Minimum-Mean-Square-Error}
\nomenclature[A]{MMSE-FDE}{MMSE Frequency-Domain Equalization}
\nomenclature[A]{MB‑DFT‑S‑OFDM‑IM}{Multi-Bands DFT‑S‑OFDM combined with {IM}}
\nomenclature[A]{NN}{Neural Network}
\nomenclature[A]{NTN}{Non-Terrestrial Network}
\nomenclature[A]{OFDM}{Orthogonal Frequency Division Multiplexing}
\nomenclature[A]{OOB}{Out-of-Band}
\nomenclature[A]{O-OFDMNet}{Optical {OFDM} Network}
\nomenclature[A]{OTFS}{Orthogonal Time Frequency Space}
\nomenclature[A]{OTFS-I/Q-IM}{{OTFS}-based In-Phase and Quadrature-IM}
\nomenclature[A]{PAPR}{Peak to Average Power Ratio}
\nomenclature[A]{PDF}{Probability Density Function}
\nomenclature[A]{QAM}{Quadrature Amplitude Modulation}
\nomenclature[A]{ReLU}{Rectified Linear Unit}
\nomenclature[A]{RF}{Radio Frequency}
\nomenclature[A]{SCS}{Subcarrier Spacing}
\nomenclature[A]{SD}{Soft-Decision}
\nomenclature[A]{SE}{Spectral Efficiency}
\nomenclature[A]{SFFT}{Symplectic Fourier Transforms}
\nomenclature[A]{SISO}{Soft-Input Soft-Output}
\nomenclature[A]{TD}{Time Domain}
\nomenclature[A]{TDL}{Time Delay Line}
\nomenclature[A]{TF}{Time-Frequency}

\nomenclature[B,01]{$M$}{number of subcarrier spacing/delay bins in DD domain}
\nomenclature[B,02]{$N$}{number of consecutive symbols/Doppler bins in DD domain}
\nomenclature[B,03]{$\Delta f$}{subcarrier spacing}
\nomenclature[B,04]{$T$}{symbol duration, $T = 1/\Delta f$}
\nomenclature[B,05]{$B$}{system bandwidth, $B = M\Delta f$}
\nomenclature[B,06]{$n$}{OTFS symbol(time) index}
\nomenclature[B,07]{$m$}{subcarrier index of FD symbol/Sample index in TD signal}
\nomenclature[B,08]{$k$}{Doppler index in the DD domain}
\nomenclature[B,09]{$l$}{delay index in the DD domain}
\nomenclature[B,091]{$P$}{number of resolvable propagation paths in the DD domain}
\nomenclature[B,092]{$L$}{number of TDL taps}

\nomenclature[B,25]{$\tilde{h}_p$}{complex DD-domain gain of the $p$-th path}
\nomenclature[B,26]{$\tau_p$}{delay of the $p$-th path}
\nomenclature[B,27]{$\vartheta_p$}{Doppler frequency of the $p$-th path}
\nomenclature[B,28]{$l_p$}{delay index of the $p$-th path}
\nomenclature[B,29]{$k_p$}{Doppler index of the $p$-th path}

\nomenclature[B,291]{$\tilde{h}_p$}{time-invariant DD-domain channel gain of the $p$-th path}
\nomenclature[B,292]{$h_{n,m,l}$}{channel coefficient of the $l$-th delay tap in TD}

\nomenclature[B,301]{$n^{'}(h)$}{atmospheric refractive index as a function of altitude $h$}
\nomenclature[B,302]{$h^{'}$}{altitude above the Earth's surface}
\nomenclature[B,303]{$N_0'$}{surface refractivity constant}
\nomenclature[B,304]{$h_0$}{scale height of the atmosphere}

\nomenclature[B,305]{$d_{rf}$}{refracted propagation path length}
\nomenclature[B,306]{$R$}{Earth radius}
\nomenclature[B,307]{$H'$}{satellite orbital altitude}
\nomenclature[B,309]{$\theta$}{elevation angle}
\nomenclature[B,311]{$\kappa_i$}{Chebyshev--Gauss quadrature node}
\nomenclature[B,312]{$\omega_i$}{Chebyshev--Gauss quadrature weight}
\nomenclature[B,313]{$Q$}{number of Chebyshev--Gauss quadrature points}

\nomenclature[B,314]{$\mathcal{P}_{PL}$}{large-scale path loss gain}
\nomenclature[B,315]{$c$}{speed of light}
\nomenclature[B,316]{$f_c$}{carrier frequency}
\nomenclature[B,317]{$\alpha_p$}{path loss exponent}
\nomenclature[B,320]{$m'$}{Nakagami-$m$ fading shape parameter}
\nomenclature[B,321]{$K_{Sct}$}{scattered (nLoS) component power parameter}
\nomenclature[B,322]{$K_{LoS}$}{line-of-sight (LoS) component power parameter}
\nomenclature[B,323]{$b_0$}{half of the nLoS average received power}
\nomenclature[B,324]{$\Omega$}{average received power of the LoS component}

\nomenclature[B,325]{$\mathcal{P}_{abs}$}{atmospheric absorption (transmittance)}
\nomenclature[B,326]{$\tau_i$}{optical thickness of the $i$-th atmospheric gas}

\nomenclature[B,327]{$\psi(t)$}{Earth-centered angular displacement at time $t$}
\nomenclature[B,329]{$\omega_{R,u}(t)$}{relative angular velocity between satellite and user}

\nomenclature[B,342]{$P_s$}{transmit power}
\nomenclature[B,3515]{$G$}{number of index-modulation subgroups}
\nomenclature[B,3517]{$D$}{size of each OTFS-IM subblock, $D=MN/G$}
\nomenclature[B,3518]{$b_g$}{number of bits in the $g$-th subgroup}
\nomenclature[B,3519]{$b_g^1$}{index bits in the $g$-th subgroup}
\nomenclature[B,3520]{$b_g^2$}{modulation bits in the $g$-th subgroup}

\nomenclature[B,3521]{$\mathcal{Q}$}{modulation constellation size}
\nomenclature[B,3523]{$\boldsymbol{x}_g$}{IM symbol vector of the $g$-th subgroup}
\nomenclature[B,3528]{$\mathbf{y}$}{received signal vector in TD}
\nomenclature[B,3529]{$\mathbf{s}$}{transmitted signal vector in TD}
\nomenclature[B,3530]{$\mathbf{v}$}{AWGN vector}
\nomenclature[B,35301]{$\tilde{\mathbf{z}}$}{MMSE-estimated DD-domain symbol vector}

\nomenclature[B,3531]{$\mathbf{H}$}{channel matrix in TD}

\nomenclature[B,3533]{$\tilde{\mathbf{Y}}$}{received signal matrix in the DD domain}

\nomenclature[B,3537]{$\tilde{\mathbf{H}}$}{channel matrix in DD domain}

\nomenclature[B,3599]{$\mathbf{S}$}{OTFS signal matrix in TD}
\nomenclature[B,5507]{$f_g(\cdot)$}{encoder function of the $g$-th band}
\nomenclature[B,5508]{$g_g(\cdot)$}{decoder function of the $g$-th band}

\nomenclature[B,5518]{$\Pi$}{interleaver operator}
\nomenclature[B,5519]{$\Pi^{-1}$}{de-interleaver operator}

\nomenclature[B,5520]{$\mathbb{H}$}{equivalent fading channel operator}
\nomenclature[B,5521]{$H_e$}{effective channel matrix in the DD domain}
\nomenclature[B,5526]{$\boldsymbol{W}^{(f/g)}$}{weight  of the  encoder/decoder}
\nomenclature[B,5527]{$\boldsymbol{b}^{(f/g)}$}{bias of the  encoder/decoder}

\nomenclature[B,5545]{$L(b)$}{log-likelihood ratio of coded bit $b$}

\printnomenclature

% \fi

\section{INTRODUCTION}
\IEEEPARstart{T}{he} next decade is expected to witness a further surge in wireless applications, such as navigation, underwater communication, and ubiquitous suburban coverage, that demand seamless connectivity across diverse environments \cite{Zhu_Creating_2022_Aug}\cite{Xu_Sixty_2019}. 
Non-Terrestrial Networks ({NTN}s), encompassing satellites, high-altitude platforms, and uncrewed aerial systems are emerging as the main enablers of the next generation of ubiquitous connectivity~\cite{Ma_Satellite_2024_Feb}. Given the 3GPP’s integration of {NTN} support into their Release‑17 and beyond, there is a paradigm shift in how channels are characterized, modeled, and optimized in {NTN} environments\cite{ntnoverview}. {NTN}s unlock access for remote, rural, maritime, aerial, and disaster‑affected regions, addressing persistent gaps in terrestrial coverage and advancing global digital inclusion\cite{ntn2}. However, as 5G‑{NTN} evolves into 6G, with applications like high‑data‑rate THz links and integrated sensing and communication, the accurate characterization of the channel models becomes more strict and crucial, especially for Low Earth Orbit ({LEO}) satellite-ground communication when considering the high Doppler variability due to satellite motion, longer latency due to the increased path length, atmospheric effects, and distinctive large-scale fading profiles~\cite{leooverview}. Exploring these unique {NTN} channel characteristics is essential to ensure robust, reliable, and low‑latency communications across heterogeneous 3D networks to guide the next‑generation standardization and transceiver designs for all terrestrial and non‑terrestrial users.
When the Line-of-Sight ({LoS}) reception from terrestrial base stations is blocked, {LEO} satellite constellations offering robust {LoS} links might be harnessed~\cite{Liu_Space_2018_4quarter}. However, {LEO} downlinks  face unique challenges, including: 1) severe Doppler shifts owing to the satellites' velocity nearing 7.8 km/s, 2) molecular absorption above 20 GHz, particularly from the water vapor and oxygen, 3) atmospheric refraction that bends propagation trajectory, 4) small-scale fading due to atmospheric turbulence imposed by temperature and pressure  fluctuations, and 5) weather-induced attenuation from rain, fog, and clouds. These effects render {LEO} channels highly doubly dispersive~\cite{dobulyntnchannel}.\par
As such, Orthogonal Time Frequency Space ({OTFS}) modulation emerges as a compelling solution, owing to its robustness against Time-Frequency ({TF}) selectivity~\cite{otfs1,otfsmag}. Unlike Orthogonal Frequency Division Multiplexing (OFDM), which operates in the {TF} domain, {OTFS} modulates information in the Delay-Doppler ({DD}) domain. This unique domain transformation enables {OTFS} to fully exploit the channel's diversity and significantly enhance the resilience against Doppler shifts and multi-path delay spread. As a result, {OTFS} provides superior reliability and Spectral Efficiency ({SE}) in rapidly time-varying channels, making it a key enabler for future wireless systems~\cite{otfsntn}.\par

Despite these advantages, {OTFS}-based systems still face challenges such as its high Peak-to-Average Ratio ({PAPR}), which causes non-linear distortion in power amplifiers and deteriorates its Bit Error Rate ({BER}) performance. Conventional {PAPR} mitigation techniques, such as clipping as well as Iterative Clipping and Filtering, often introduce signal distortion, which perturbs {OTFS}'s unique signal structure \cite{otfssmae}\cite{ jddimae}.   In \cite{otfsprecoderguide}, Yusuf \textit{et al.} proposed transmit precoder-based {PAPR} reduction, which employs classical orthogonal transform-based precoders, such as Zadoff-Chu transform, Walsh-Hadamard transform, and discrete cosine transform-based techniques, which can significantly reduce the {PAPR} with only a minor compromise in {BER} performance. Then, Praksh \textit{et al.}\cite{otfsprecoderamp} introduced an iterative amplitude-domain precoder, which reduces the {PAPR} by about 5 dB, while maintaining a {BER} performance close to that of the standard {OTFS} \cite{otfsprecoderamp}.\par
In parallel, Index Modulation ({IM}) techniques have gained significant attention as a benefit of improving both the {SE} and Energy Efficiency ({EE}) by conveying information not only through conventional modulation symbols but also through the activated/deactivated indices of transmission entities, such as subcarriers, antennas, or {DD} domain indices~\cite{imconcept,mim,jmim}. Integrating {IM} into {OTFS}, termed as {OTFS-IM} \cite{otfsim1}, further enhances the system's performance by leveraging the sparsity and diversity in the {DD} plane. In {OTFS-IM}, certain entities of {DD} domain indices are intentionally left as inactive ones, hence implicitly conveying information via the activated indices. This sparsity in the activated indices leads to inherently lower instantaneous peaks, thus improving the {PAPR} performance compared to standard {OTFS} \cite{otfsim2}. Furthermore, the {OTFS}-based In-Phase and Quadrature-IM ({OTFS-I/Q-IM}) components can attain lower {PAPR} than {OTFS-IM} schemes~\cite{otfsim3}. From a broader perspective, IM could be viewed as an enabling mechanism that can be integrated with other waveform-level or precoding techniques.
\par

 \begin{table*}[!t]
  \centering
  \caption{Contrasting our contributions to the literature: {OTFS-IM} Autoencoder in {NTN} scheme.}
  \scalebox{0.88}{
  \begin{tabular}{|l||c||c|c|c|c|c|c|c|c|c|c|c|c|} 
  \hline
  Contribution&proposed*&\cite{otfsdl}&\cite{ofdmae}&\cite{otfsper}&\cite{otfsmetalearning}&\cite{otfsae}&\cite{otfsimae}&\cite{aechannelcoding}&\cite{ddofdm}&\cite{otfsturbo}&\cite{turbo}&\cite{otfsntn}&\cite{dftmb}\\
  \hline
  \hline
   Autoencoder structure&\checkmark&&\checkmark &\checkmark &\checkmark &\checkmark &\checkmark  &\checkmark &\checkmark &\checkmark &&&\\
  \hline
  {OTFS} system&\checkmark&\checkmark & &\checkmark &\checkmark &\checkmark& \checkmark& \checkmark& & \checkmark& & \checkmark &\\
   \hline
  {PAPR} Reduction&\checkmark& &\checkmark& &&\checkmark  & & &\checkmark &   & &&\checkmark\\
  \hline
  Channel Coding&\checkmark& &&& &&&\checkmark &\checkmark &\checkmark &\checkmark &&\\
  \hline
  {IM}&\checkmark& &&& &&&\checkmark &\checkmark &\checkmark &\checkmark &&\\
  \hline
  SD-DNN&\checkmark & & && & & & & &\checkmark & \checkmark & & \\
    \hline
  
  \textbf{MB-DFT-S structure}&\CheckmarkBold& &&& &&&& & &&&\checkmark \\
  \hline
    \textbf{{NTN} channel modeling}&\CheckmarkBold& & & && & & & &&&\checkmark& \\
   \hline
   
  \end{tabular}
  }
  
  \label{Table:ref}
\end{table*}
For instance, the Discrete Fourier Transform ({DFT}) Spreading-{OFDM} ({DFT-S-OFDM}) emulates single-carrier transmission that exhibits low {PAPR}, which is standardized in the 4G, Long-Term-Evolution (LTE), and 5G New Radio~\cite{Xu_Sixty_2019}. However, the {DFT} spreading results in the loss of bandwidth efficiency due to the inevitable excess bandwidth of the pulse shaping filter harnessed for the Out-of-Band ({OOB}) components' suppression~\cite{scfdmaup}. A compelling alternative is Multi-Band ({MB}) DFT‑S‑OFDM combined with {IM} (MB‑DFT‑S‑OFDM‑IM) that intrinsically amalgamates multi-carrier transmission and {IM}, mitigating {OFDM}’s flaws without sacrificing its advantages. Specifically, DFT-based pre-coding reduces the {PAPR}, while {IM} attains a throughput  exceeding that of conventional {OFDM}. The design inherently achieves frequency diversity as a benefit of its uniform energy distribution across subcarriers, which also guards against Carrier Frequency Offsets ({CFO}). Furthermore, {OOB} filtering is applied in the Time Domain ({TD}) before {DFT}, which preserves the subcarrier orthogonality~\cite{dftmb}. In a nutshell, combining {DFT-S} with {OTFS} improves resilience against {CFO}, fractional Doppler shifts, and interference~\cite{dftsotfs}.

Recent advances in Deep Learning ({DL}) have significantly enhanced {OFDM} and {OTFS} systems by addressing challenges, such as Channel Estimation ({CE}), signal detection, and {PAPR} reduction~\cite{dnnphy,otfsdl}. In~\cite{ofdmae}, Kim and Lee conceived the method of using Deep Neural Network ({DNN}) Autoencoder ({AE}) architectures for reducing the {PAPR} in classical {OFDM} settings. With the aid of Soft-Decision ({SD}) detection, these {AEs} can output symbol reliability metrics instead of Hard-Decision ({HD}) for enhancing the decoding performance~\cite{hanzo1}. Compared to the aforementioned method, Sebastian \textit{et al.} proposed a more general structure that jointly treats the transmitter and receiver by Neural Networks (NNs) and  amalgamates the mapping and demapping optimizer, channel coding and decoding for transmission over arbitrary channels. In~\cite{ddofdm}, a SD-aided integrated autoencoder architecture, termed as the Optical {OFDM} Network (O-OFDMNet), has been designed for Radio Frequency (RF)-OFDM and optical {OFDM} in order to achieve improved coded {BER} performance and reduced {PAPR} at moderate complexity.

Furthermore, Tek \textit{et al.}~\cite{otfssmae} introduced a {DNN} autoencoder framework of {OTFS} that jointly learns encoder/decoder mappings for compressing the modulated signal peaks and reconstruct signals, achieving {PAPR} reduction without eroding the {BER} performance under nonlinear distortions. Similarly, Autoencoder-based Enhanced {OTFS} ({AEE-OTFS}) is proposed in \cite{otfspaprae}, exploring the end-to-end learning benefits across both of the mapping and demapping functions of {OTFS}, demonstrating the improvement of {SE}, {EE}, and robustness under imperfect Channel State Information ({CSI}). As a further advance,  Tek and Basar proposed Joint {DD}-{IM} with OTFS (JDDIM‑OTFS)~\cite{jddimae}, activating specific delay or Doppler resource bins to enhance the performance under doubly dispersive channels. Additionally, the Autoencoder‑Enhanced JDDIM‑OTFS (AEE‑JDDIM‑OTFS) framework~\cite{jddimae} exploits DNN autoencoders to jointly optimize the mapping and demapping, thus striking an attractive trade-off between the enhancement of {EE}/{SE} and {PAPR} reduction.

Against the above backdrop, the novel contributions of this paper are contrasted to the state-of-the-art in Table \ref{Table:ref}, which are further elaborated on as follows:
\begin{itemize}
 \item We propose a novel {AE}-based framework for OTFS‑IM systems that jointly considers {PAPR} reduction and detection. Our architecture directly integrates {OTFS} modulation and demodulation into the {AE} design. The encoder is trained for minimizing the {PAPR} of the transmitted waveform, while the soft-decoder reliably reconstructs the original information at the receiver. Our specifically designed loss function strikes a compelling trade-off between {PAPR} reduction and {BER}, carefully tuned via a selected hyperparameter. Additionally, a two-step training strategy is conceived for enhanced convergence stability in unison with our {SD} decoder.
 
 \item  We develop a {DD}-domain OTFS-based Generalized Subcarrier IM {(OTFS-GFIM)} architecture employing an MB-DFT-S-based AE capable of supporting both {HD} and {SD} detection, where the encoder jointly selects the active {DD} indices and performs constellation shaping. Firstly, by applying DFT-S pre-coding across subcarriers in the {OTFS-IM} scheme,  the {PAPR} is significantly reduced, since the signal energy is more uniformly distributed across all {DD} bins. Secondly, the {MB} design enhances the flexibility of the {OTFS-IM} autoencoder in terms of its performance, orchestrating a tunable trade-off among the throughput, {BER} performance, and {PAPR} by adjusting the size of the individual {DD} bands. The performance is then further improved at a moderate detection complexity with the aid of channel coding techniques. The autodecoder is responsible either for reconstructing the transmitted symbols or for calculating the Log Likelihood Ratios ({LLR}s) under both {HD} or {SD} detection  at the receiver, respectively, enabling robust detection under hostile fading conditions.

 \item To accurately characterize the challenging propagation environment in {NTN}s, we introduce a comprehensive downlink channel model for {LEO} satellite links to ground terminals, based on \cite{otfspaprae}. Our model incorporates the rotational dynamics of the Earth, embedding the Earth's angular velocity into the relative motion calculation between {LEO}s and users. 
 In the simulations, we rigorously characterize the key fading-channel characteristics under time-varying {NTN} scenarios, including different elevation angles, {LoS}/non-Line-of-Sight ({nLoS}) conditions, and Doppler spreads, to verify that our small-scale fading model matches the practical by observed behavior. Our results show that the proposed system exhibits robustness to practical {NTN} channels, confirming its suitability for {LEO}-based {NTN} links.
\end{itemize}
The rest of this paper is organized as follows. In Section II, we describe the {OTFS} system and the {NTN} channel model, followed by the MB-DFT-S and {IM} structure. The proposed  AE-based architecture is presented in Section III. Our performance analysis and simulation results are discussed in Section IV, and finally, we conclude in Section V.
\section{SYSTEM MODEL}
In this section, we first review the {OTFS} architecture and its channel model. Then, we extend our discussions to the {NTN} channel characteristics. Finally, the intricate principles of the proposed MB-DFT-S-OTFS-IM modulation/demodulation scheme are presented, together with its {PAPR} analysis.
\begin{figure}[!htb]
\centering
{\includegraphics[width=8cm]{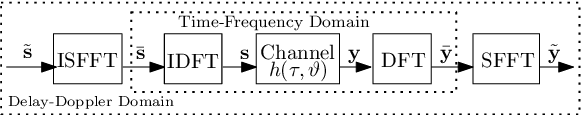}}%
\caption{Block diagram of the {OTFS} system.}
\label{fig:otfs}
\end{figure}
\subsection{{OTFS} System}

At the transmitter, the binary information bits are first grouped and mapped onto amplitude phase modulation symbols, which are arranged on a two-dimensional Delay–Doppler ({DD}) grid. As illustrated in Fig.~\ref{fig:otfs}, these {DD}-domain symbols are then transformed to the Time–Frequency ({TF}) domain via the Inverse Symplectic Finite Fourier Transform ({ISFFT}). Then an Inverse Discrete Fourier Transform ({IDFT}) is applied across each {OFDM} symbol to generate the corresponding time-domain samples. In this way, {OTFS} effectively realizes a pair of 2D transforms at the transmitter and receiver, representing information symbols in the {DD} domain rather than directly in the {TF} domain of conventional {OFDM} systems.
We consider an {OTFS} system having $M$ subcarriers and $N$ symbols spreading over $N$ time slots, where the Subcarrier Spacing ({SCS}) and symbol duration are denoted as $\Delta f$ and $T$, respectively. Hence, the transmitted signal frame has a total duration of $NT$ seconds and its bandwidth of $B = M/T = M\Delta f$ Hz.
An {OTFS} frame may be viewed as applying {ISFFT}-based pre-coding to generate \(N\) consecutive {OFDM} symbols, each having \(M\) subcarriers.
\begin{table}[!htb]
\caption{{OTFS} Notations for Each Domain.}

  \centering
  \scalebox{0.9}{
  \begin{tabular}{|c|l|c|l|} 
  \hline
 &TD&FD&{DD}\\
  \hline
  Transmitter& $s_{n,m}$&$\bar{s}_{n,m}$& $\tilde{s}{[k,l]}$\\
  \hline
  Channel& $h_{n,m,l}$&$\bar{h}_{n,m}$&$\tilde{h}_{p}\omega^{k_{p}(nM+m-l_p)}_{MN}$\\
  \hline
  Receiver& $s_{n,m}$&$\bar{y}_{n,m}$&$\tilde{y}{[k,l]}$\\
  \hline
  \end{tabular}
  \label{Table:otfs}
  }
\end{table}
Additionally, we consider a single antenna at the transmitter and receiver of our system. The TD, FD, and {DD} notations are shown in Table~\ref{Table:otfs}.
subca\subsubsection{General Modeling}
 Generally, under a doubly-selective fading channel, the TD received signal can be modeled as\cite{otfs1,otfsntn}
\begin{align}\label{eq:otfs}
y(t) = & \iint \tilde{h}(\tau,\vartheta)s(t-\tau) e^{j2\pi \vartheta (t-\tau)}d\tau d\vartheta\notag\\
&+ v(t)|_{t=\frac{nT}{M}=\frac{n}{M\Delta f}},
\end{align}
where $\tau$ and $\vartheta$ represent the delay and Doppler frequency, while
$s(t)$ and $v(t)$ denote the TD  signal and Additive White Gaussian Noise ({AWGN}). The corresponding symbol period is $T = (1/[\Delta f])$, where $\Delta f$ refers to the SCS.
Then, the {DD}-domain fading gain in (\ref{eq:otfs}) is expressed as
\begin{align}\label{eq:otfsading}
\tilde{h}(\tau,\vartheta) = \sum_{p=0}^{P-1} \tilde{h}_p \delta(\tau-\tau_p)\delta(\vartheta-\vartheta_p)|_{\tau_p=\frac{l_p}{M\Delta f},\vartheta_p=\frac{k_p}{NT}}.
\end{align}
Here we have $\tau_{max} >(T/M)$, resulting in frequency selectivity. There are $P$ resolvable paths in the {DD}-domain falling into $L$ Time Delay Line ({TDL}) taps in the {TD}, characterized by $\tau_p=([l_pT]/M)=([l+p]/[M\Delta f])$ with $l_p\in[0,L-1]$. Then, when the maximum Doppler frequency $f_D$ becomes comparable to the {SCS} $\Delta f$, time-selectivity is encountered.  The components of Doppler frequencies are all substantially lower than the overall bandwidth, mathematically formulated as $\{\vartheta_p << M\Delta f\}^{P-1}_{p=0}$. Hence, the fading gain, denoted as $\tilde{h}_pe^{([j2\pi \vartheta_pn]/[M\Delta f])}$, remains nearly fixed over a sampling period of $(T/M)$, while varies for $n=0,1,\cdots,M-1$ within an {OTFS} symbol time period $T$. Based on these assumptions, the time-varying frequency-selective fading can be uniquely represented by the time-invariant parameters of fading gain $\tilde{h}_p$, Doppler index $k_p$, and delay index $l_p$, where sampling in the {DD} domain.
Furthermore, the relationship between the {TD} and {DD} fading representation maybe expressed as 
\begin{align}\label{eq:channel relation}
\tilde{h}_{n,m,l} = \sum_{p=0}^{P-1} \tilde{h}_p \omega_{MN}^{k_p(nM+m-l_p)}\Lambda(k_p,l_p)|_{l=l_p},
\end{align}
where $\omega_{MN}^{k_](nM+m-l_p)}\Lambda(k_p,l_p)=\text{exp}(j([2\pi k_p(nM+m-l_p)]/[MN]))$ and $h_{n,m,l}$ models the $l$-th {TDL} tap for the $m$-th sample in the $n$-th {OFDM} symbol, while $\Lambda(k_p,l_p)=1|_{l=l_p}$ or $\Lambda(k_p,l_p)=0|_{l=l_p}$ when the $p$-th {DD}-domain resolvable path falls within the  $l$-th {TDL} tap, or otherwise, respectively. Then the separate propagation paths, which experience the same delay, obtain a higher degree of freedom in channel modeling. With the aid of an appropriate windowing pair at the transmitter and receiver, the fractional delay and Doppler indices of the off-the-grid paths can be mitigated\cite{otfsdev}. Then we can exploit this relationship within the {NTN} channel model.\par
\subsubsection{{OTFS} Modulation}
All the subblocks are passed through the {OTFS} block generator for generating the {DD} domain signal as a {DD} domain matrix $\tilde{\textbf{S}}\in\mathbb{C}^{M\times N}$ and for each grid, we have $\{\{{\tilde s}[k,l]\}_{k=0}^{N-1}\}_{l=0}^{M-1}$ with size $M \times N$, which is shown in Fig.~\ref{fig:sturcture}.

The {OTFS} block creator can generate a vectorized {DD} domain symbol, formulated as 
\begin{align}\label{DDvecsignal}
\tilde{\textbf{s}}=vec(\tilde{\textbf{S}})=[{\tilde s}[0,0],\cdots,{\tilde s}[M-1,N-1]]^T.
\end{align}
For an {OTFS} frame, each block $\tilde{\textbf{S}}$ in the {DD} domain is first converted into the {TF} domain through the 2D ISFFT. The scalar form of this operation can be expressed as
\begin{align}\label{ddisfft}
{{\bar s}[n,{\bar m}]} = { {\frac{1}{{\sqrt{NM}}}} }\sum_{k=0}^{N-1 }\sum_{l=0}^{M-1}{\tilde s}[k,l]\omega_N^{nk}\omega_M^{-{\bar m}l}.
\end{align}

Equivalently, the {ISFFT} operation can be expressed in a compact matrix form as
\begin{align}\label{ddisfft_matrix}
\bar{\textbf{S}} = \textbf{F}_M \tilde{\textbf{S}} \textbf{F}_N^H,
\end{align}
where $\textbf{F}_M \in \mathbb{C}^{M\times M}$ and $\textbf{F}_N \in \mathbb{C}^{N\times N}$ denote the normalized Discrete Fourier Transform (DFT) matrices, and $(\cdot)^H$ represents the Hermitian transpose.

As shown in Fig.\ref{fig:otfs}, we consider an ideal bi-orthogonal waveform, which is based on the practical rectangular waveforms in \cite{otfsofdm1,otfsofdm2,otfsofdm3}. More specifically, following the {ISFFT}, the transmitter performs {IDFT} in the {TD} as ${\textbf{S}}=\bar{\textbf{S}}\textbf{F}_N^H=\textbf{F}_M\tilde{\textbf{S}}$, which is shown in Fig.~\ref{fig:sturcture}. Similarly, we can have
\begin{align}\label{ddidft}
{{s}[n, m}] = { {\frac{1}{{\sqrt{M}}}} }\sum_{{\bar m}=0}^{M-1}{\tilde s}[n,{\bar m}]\omega_M^{m{\bar m}}={\frac{1}{{\sqrt{N}}}}\sum_{k=0}^{N-1}{\tilde s}[k,m]\omega_N^{nk}.
\end{align}
After transmission over a doubly selective channel, the continuous-time received signal spanning $N$ {OFDM} symbol durations is modeled as
\begin{align}\label{eq:otfs2}
y(t) =& \iint \sum_{p=0}^{P-1} \tilde{h}_p \delta(\tau-\tau_p)\delta(\vartheta-\vartheta_p) s(t-\tau)\notag\\
&\times e^{j2\pi \vartheta (t-\tau)}d\tau d\vartheta + v(t)|_{t=\frac{nM+m}{M}T,\tau=\frac{l}{M\Delta f},\vartheta=\frac{k}{NT}}.
\end{align}

To simplify the equation, we can express the {TD} signal of ($\ref{eq:output relation}$) in discrete time as 
\begin{align}\label{eq:rect}
{{y}[n, m}] = \sum_{p=0}^{P-1}{\tilde h_p}\omega_{MN}^{k_p[nM+m-l_p]}s[n,\langle m-l_p\rangle_M]+v[n,m].
\end{align}
By vectorizing the received samples, the {TD} input-output relationship is compactly written in matrix form as
\begin{align}\label{recematrix}
\textbf{y} = \textbf{H}\textbf{s} + \textbf{v},
\end{align}
where $\textbf{y}, \textbf{s}, \textbf{v} \in \mathbb{C}^{MN\times 1}$, and $\textbf{H} \in \mathbb{C}^{MN\times MN}$ is the {TD} channel matrix. At the receiver side, the signal is reshaped into an $M\times N$ matrix $\textbf{Y}$. Applying the {DFT} and {SFFT} successively, the {DD} domain received signal matrix is obtained as
\begin{align}\label{isfftmatrix}
\tilde{\textbf{Y}} = \textbf{F}_M \bar{\textbf{Y}} \textbf{F}_N^H = \textbf{Y}\textbf{F}_N^H.
\end{align}

Based on this, applying column-wise vectorization yields the end-to-end {OTFS} input-output relationship:
\begin{align}\label{eq:oimatrix}
\tilde{\textbf{y}} = \tilde{\textbf{H}}\tilde{\textbf{s}} + \tilde{\textbf{v}}.
\end{align}
Here, the $i$-th element of $\tilde{\textbf{y}} \in \mathbb{C}^{MN\times 1}$ maps to the {DD} grid as $\tilde{\textbf{y}}_i = \tilde{y}[l,k]$, where the column index is $k = \lfloor i/M \rfloor$ and the row index is $l = i - kM$. The elements of $\tilde{\textbf{s}}$ and $\tilde{\textbf{v}}$ are mapped similarly. The {DD}-domain effective channel matrix $\tilde{\textbf{H}} \in \mathbb{C}^{MN\times MN}$ is sparse and time-invariant, with its nonzero elements given by $\tilde{\textbf{H}}_{i,j} = \tilde{h}_p \tilde{T}(k,l,k_p,l_p)$ associated with $j = M \langle k \rangle_N + \langle l \rangle_M$. 

Finally, utilizing the vectorized equivalent model in \eqref{eq:oimatrix}, the Minimum Mean Square Error ({MMSE}) detector \cite{mmse} is formulated as
\begin{align}\label{eq:mmse}
\tilde{\textbf{z}} = (\tilde{\textbf{H}}^H\tilde{\textbf{H}} + N_0\textbf{I}_{MN})^{-1}\tilde{\textbf{H}}^H\tilde{\textbf{y}}.
\end{align}

\subsection{{NTN} Channel Modeling}
\subsubsection{Large-Scale Path Loss and Small-Scale Fading}
Due to the atmospheric refractivity, a Snell-law-based path loss model is exploited.
Given a long trajectory under the effect of atmospheric refraction, the electromagnetic waves transmitted from the satellites follow a curved path instead of a straight {LoS} link. Given by the ITU-R recommendation sectors \cite{ntnchannel} with ${N_0^{'}} = 315 \times 10^{-6}$ and $h^{'}_0 = 7.5$ km, the atmospheric refractive index is formulated as $ n^{'}\left( h \right)=  1 + {N_0^{'}}\exp \left( {-\frac{h}{{{h_0}}}} \right)$. By considering the atmospheric refraction, the path-length is calculated as \cite{ntnchannel}:
\begin{align} \label{speed}
{d_{rf}}\left( h^{'} \right) =& \sum\limits_{i = 1}^Q {\frac{{H'{\omega _i}}n^{'}\left( {{\kappa _i}} \right)}{2}{{\left( {1 - {{\cos }^2}\left( {\frac{2i-1}{{2Q }}\pi } \right)} \right)}^{ \frac{1}{2}}}} \notag\\
&\times {\left( {1 - {{\left( {\frac{{{n^{'}_0}\cos \left( {{\theta _0}} \right)}}{{n^{'}\left( {{\kappa _i}} \right)\left( {1 + \frac{{{\kappa _i}}}{R}} \right)}}} \right)}^2}} \right)^{ - \frac{1}{2}}},
\end{align}
where we denote the Earth's radius as $R$, the altitude of the satellite's orbit as $H'$, and the initial (detected) elevation angle as $\theta_0$. The following settings are formulated as ${\kappa _i} = \frac{{H'\left( {\cos \left( {\frac{{2i - 1}}{{2Q}}\pi } \right) + 1} \right)}}{2}$, ${\omega _i} = \frac{\pi }{Q}$, and $Q=100$ for computing the Chebyshev-Gauss quadrature~(The Gauss quadrature is introduced in the Section 4.6 of \cite{Press2007Numerical}).   

Subsequently, the path loss is formulated as:
\begin{align}\label{pathloss}
{{\cal P}_{PL}} = {\left( {\frac{c}{{4 \pi {f_c}}}} \right)^2}d_{rf}^{ - \alpha_{p} },
\end{align}
where $c$ is the speed of light, ${{f_c}}$ is the carrier frequency, and $\alpha_p$ is the path loss exponent.\par

\begin{figure*}[!b]
\hrulefill
{\small
\begin{align}\label{Doppler}
f_{D,max} =  - \frac{{f_c}}{c}\frac{{R{H_{os}}\sin \left( {\psi \left( {t,{t_0}} \right)} \right)\cos \left( {{{\cos }^{ - 1}}\left( {\frac{{R\cos {\theta _{\max }}}}{{{H_{os}}}}} \right) - {\theta _{\max }}} \right){\omega _{R,u}(t)}}}{{\sqrt {{R^2} + H_{os}^2 - 2R{H_{os}}\cos \left( {\psi \left( {t,{t_0}} \right)} \right)\cos \left( {{{\cos }^{ - 1}}\left( {\frac{{R\cos {\theta _{\max }}}}{{{H_{os}}}}} \right) - {\theta _{\max }}} \right)} }}.
\end{align}
}
\end{figure*}

With the consideration of atmospheric turbulence, the small-scale fading obeys a Shadowed-Rician distribution~\cite{ntnchannel,Abdi2003Anew}. We define the index of the {LoS} path as $l = 0$, modeled by Nakagami-$m$ fading, while the {nLoS} component is constituted by the conventional Rayleigh scattering. The normalized Probability Density Function (PDF) and Cumulative Distribution Function (CDF) of the Shadowed-Rician distribution in the power domain are expressed as \cite{ntnchannel}:
\begin{align}\label{PDF_h_ST_K}
{f_{{{\left| {{h }} \right|}^2}}}(x) = & \sum\limits_{k = 0}^{m' - 1} \binom{m-1}{k}\frac{{{{\left( {{K_{Sct}}} \right)}^{m' - k - 1}}{{\left( {{K_{LoS}}} \right)}^k}}}{{k!{{\left( {{K_{Sct}} + {K_{LoS}}} \right)}^{m'}}}}\notag\\
&\hspace*{-1cm} \times {{\left( {\frac{x}{{{K_{Sct}} + {K_{LoS}}}}} \right)}^k}\exp \left( { - \frac{x}{{{K_{Sct}} + {K_{LoS}}}}} \right),\\
\label{CDF_h_ST_K}
{F_{{{\left| {{h }} \right|}^2}}}(x) = & 1 -  {\sum\limits_{k = 0}^{m' - 1} {\binom{m'-1}{k}} \frac{{{{\left( {{K_{Sct}}} \right)}^{m' - k - 1}}{{\left( {{K_{LoS}}} \right)}^k}}}{{{{\left( {{K_{Sct}} + {K_{LoS}}} \right)}^{m' - 1}}}}} \notag\\
& \hspace*{-1.3cm}\times \sum\limits_{p = 0}^k {\frac{x^p}{{p!}} {{{\left( {\frac{1}{{{K_{Sct}} + {K_{LoS}}}}} \right)}^p}\exp \left( { - \frac{x}{{{K_{Sct}} + {K_{LoS}}}}} \right)}} ,
\end{align}
where the shape parameter of the Nakagami-$m$ distribution is assumed to be an integer\footnote{When $m'=1$, the fading behaves exactly like Rayleigh fading, otherwise, $m'\gg1$, the fading behaves similar to Rician fading. In this paper, we consider an easy assumption that integer of $m'$ is applied in distinct different scheme to elaborate the dispersion of different channels and build the relationship in the formula\cite{ntnchannel}.}, denoted as $m$. We also have ${K_{Sct}} = 2{b_0} $ and ${K_{LoS}} = \frac{\Omega }{m'}$, where $2 b_0$ is the received power of the {nLoS} component, and $\Omega$ is the received power of the {LoS} component. Hence, the normalization is mathematically represented by an equation, formulated as $2b_0+\Omega = 1$.
\subsubsection{Atmospheric Absorption and Maximum Doppler analysis}
As for the atmospheric absorption, a simplified Beer-Lambert-law-based model for the transmittance is expressed as \cite{ntnchannel}:
\begin{align} \label{abs}
{{\cal P}_{abs}}\left( \tau_i \right) =  {\exp{\left( -\sum\nolimits_i \tau_i \right)}},
\end{align}
where $\tau_i$ is the optical thickness through the atmosphere. The value of $\tau_i$ for different carrier frequencies and gases can be found in \cite{ROTHMAN_The_2013,EODG}. For instance, if a 26 GHz subcarrier is selected, the gases with effective absorption are water vapours and Oxygen. Their optical thicknesses are 0.183 and 0.0564, respectively. Therefore, the transmittance is calculated as ${{\cal P}_{abs}}\left( \tau_i \right) = \exp(-(0.183+0.0564))=0.7871$. Hence, 78.71\% of the energy is received under 26 GHz carriers.

Let us assume that the earth is a perfect sphere and the orbits of {LEO}s are perfect circles concentric with the Earth. Note that the relative angular velocity between a {LEO} and a terrestrial user is ${\omega _{R,u}(t)}$, while the maximum elevation angle, denoted as $\theta_{max}$, is obtained at the time $t_0$. With $\psi (t)$ as the central angle reached after a time $t$ has elapsed, we have $\dot \psi (t) = \frac{{d\psi (t)}}{{dt}} = {\omega _{R,u}(t)}$, where we define $\psi \left( {t,{t_0}} \right) = \psi (t) - \psi \left( {{t_0}} \right)$. If the distance from the Earth's center to the {LEO} is formulated as $H_{os} =R+H$, the maximum Doppler frequency is represented by \eqref{Doppler} \cite{ntnchannel}.

\subsubsection{{OTFS} Transmission in {NTN} Channels}
For a direct link, there are $P$ {DD}-domain resolvable paths associated with the  maximum delay of $\tau_{max}$. For a conventional Rician channel based on (\ref{eq:channel relation}), the {nLoS} paths are associated with $1\le l \le L-1$. Then the time-invariant gain $\tilde{h}_{p}$ is generated based on a complex Gaussian distribution having zero mean and a variance of $(1/[(K_{rician}+1)(P-1)])$. The {nLoS} delay indices $\{l_p\}_{p=1}^{P_{SD}-1}$ are randomly generated in the interval $\mathbb{Z} \in [0,L-1]$, where we have $L=[\tau_{max}M\Delta f]$. The Doppler indices are randomly generated over the
range of $\mathbb{R} \in [-k_{max},k_{max}]$, where the maximum index is given by $k_{max}=[f_{D}NT]$. Furthermore, the {LoS} path associated with $l=l_p=0$ and $p=0$ is given by
\begin{align}\label{eq:kfact1}
{h}_{n,m,0} = \sqrt{\frac{K_{Rician}}{K_{Rician+1}}} \omega_{MN}^{k_0(nM+m)},
\end{align}
where the {LoS} fading gain $\tilde{h}_{p}$ of (\ref{eq:channel relation}) associated with $p=0$ is simply given by $\tilde{h}_{0}=\sqrt{([K_{Rician}]/[K_{Rician+1}])}$, while the {LoS} Doppler index is given by $k_0=[f_D \cos{(\phi_0)NT]}$.
Then, based on the analysis between the Nakagami-$m$ distribution and Rician distribution, we have the relationship of $K_{Riciain}=\Omega/2b_0$. Then we have:
\begin{align}\label{eq:kfact2}
{h}_{n,m,0} = \sqrt{\Omega} \omega_{MN}^{k_0(nM+m)}.
\end{align}
To elaborate further, the  {nLoS} components follow the complex Gaussian distribution with zero mean and a variance of $(2b_0/(P-1))$. As a result, the received signal is modeled in the {TD} and {DD} domain as follow:
\begin{align}\label{eq:output relation}
y_{n,m} = \sum_{l=0}^{L-1} {h}_{n,m,l} s_{n,m-l}=\sum_{p=0}^{P-1}\tilde{h}_p\omega
_{MN}^{k_p(nM+m-l_p)}s_{n,m-l_p},
\end{align}
where we have $\Lambda(k_p,l_p)=1$ for all the $P$ resolvable paths.
\subsubsection{Received SNR}
The received {SNR} of the terrestrial user is expressed as:
\begin{align}\label{SNR}
\gamma_{SNR} = \frac{P_s {\mathcal{P}_{PL}}{\mathcal{P}_{abs}} \left| h \right|^2 }{\sigma ^2},
\end{align}
where $P_s$ is the transmit power, $\mathcal{P}_{PL}$ is the path loss gain given in \eqref{pathloss}, $\mathcal{P}_{abs}$ is the atmospheric transmittance given in \eqref{abs}, $h$ is the Shadowed-Rician fading whose power-domain distribution is given by \eqref{PDF_h_ST_K} and \eqref{CDF_h_ST_K}, and ${\sigma ^2}$ is the variance of the {AWGN}. The Doppler in \eqref{Doppler} is mitigated by {OTFS} modulation and the {BER} performance is optimized by the channel coding used, which are investigated in the following section.

\begin{figure*}[!t]
\centering
{\includegraphics[width=13.8cm]{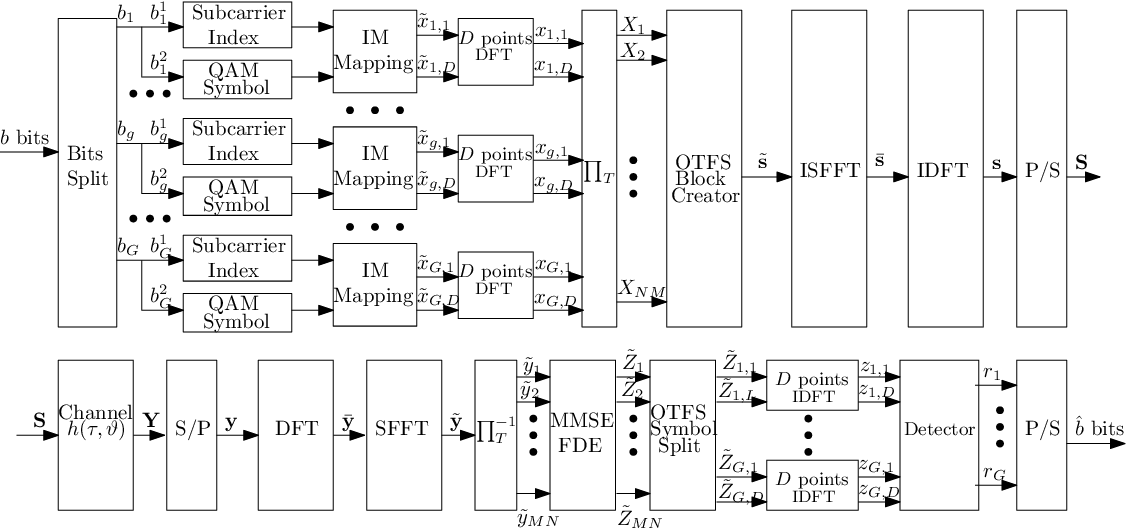}}%
\caption{Block diagram of the proposed MB-DFT-S-OTFS-IM system.}
\label{fig:sturcture}
\end{figure*}
\subsection{MB-DFT-S-OTFS-IM System}
The {DFT-S-OFDM-IM} scheme proposed is capable of achieving higher throughput than conventional {OFDM} as a benefit of our {IM} design. The DFT-spreading pre-coding operations introduce additional frequency diversity gains, which result in improved {BER} performance compared to {DFT-S-OFDM} and {OFDM-IM} using pre-coding schemes.

The {MB-DFT-S-OFDM-IM} framework further enhances flexibility by facilitating a tunable trade-off between {PAPR}, throughput, and complexity \cite{dftmb}. It can achieve a low adjustable {PAPR}, while retaining subcarrier orthogonality and supporting single- or reduced-{RF}-chain based Multiple-Input-Multiple-Output ({MIMO}) transmission. However, similar to {OFDM-IM}, {MB-DFT-S-OFDM-IM} involves extra {IM} detection complexity. Nonetheless, the {MB} arrangement mitigates this issue by reducing the search space as the number of subgroups $G$ increases. Furthermore, the number of {IM} bits can be flexibly configured to balance the performance \textit{vs.} complexity.
In contrast to {OFDM-IM}, in {OTFS-IM} the index information is carried in the {DD} domain, which is then transformed into the {TF} domain through a pair of  {SFFT}. In the following, we will detail the {OTFS-IM} and how {MB-DFT-S} can be harnessed in the {OTFS-IM} system conceived.
\subsubsection{MB-OTFS-IM(OTFS-GFIM) Mapping}
As illustrated in Fig.~\ref{fig:sturcture}, we assume a total of $b$ bits to be transmitted. To increase the {SE}, these $b$ bits are split into $G$ groups with $b_g$ bits in each group. Accordingly, an {OTFS} subblock of size $D=NM/G$ with $b_g$  $(g=1,2,\cdots, G)$ bits\footnote{In this case, we consider $G=M$ to simplify the {IM} mapping and detection. When $G< M$, each sub-block covers more than one {IM} symbol combination, increasing the number of possible activation patterns and thus introduce extra detection complexity. Conversely, when $G> M$, the detection complexity per group may reduce, while the overhead associated with managing multiple groups increase.} can be generated, which is divided into $b_g^1$ bits and $b_g^2$ bits,  corresponding to the index bits and amplitude phase modulation information bits, respectively. \par
Then, the $b_g^1$ bits are used for index selection. Then the index information of the $g$-th subblock can be expressed as 
\begin{align}\label{imdomain}
&\:a\nonumber\\
&\downarrow\nonumber\\
\boldsymbol{x}_{g}=[x_{g,1}\: \cdots\: x_{g,a-1}\: &\:0 \:\cdots \:x_{g,a+1} \: \cdots\: x_{g,D-1}],
\end{align}
where the zero padding index $a$ convey $b_g^1=log_2D$ bits, while the remaining $b_g^2=(D-1)log_2\mathcal{Q}$  bits are mapped into the $\mathcal{Q}$-ary signal constellation conveyed by the active index subsets.
\subsubsection{DFT-S Arrangement}
 As shown in Fig.~\ref{fig:sturcture}, in the $g$-th group,   before mapping the information symbols onto the {DD} plane, we perform a $N$-point {DFT} operation on the data symbols along the Doppler axis, formulated as
\begin{align}
\tilde x^{DD}_g[k]=\frac{1}{\sqrt{N}}\sum_{n=0}^{N-1} x_g[n]\omega^{nk}_N,
\label{eq:dfts}
\end{align}
for $k=0,1,\cdots,N-1 $. 
By applying {DFT}-spreading along the Doppler dimension in the {DD} grid of an {OTFS} system, each {IM} symbol is distributed across multiple Doppler bins and subcarriers, which enhances frequency (and Doppler) diversity and mitigates the impact of deep fades. Simultaneously, the spreading technique reduces the instantaneous power peaks in the {TD} transmit signal, thereby reducing the {PAPR} of the signal. Thus, the {DFT-S} operation improves the {BER} performance and enhances the {EE}.

\begin{figure*}[!t]
\centering
{\includegraphics[width=14cm]{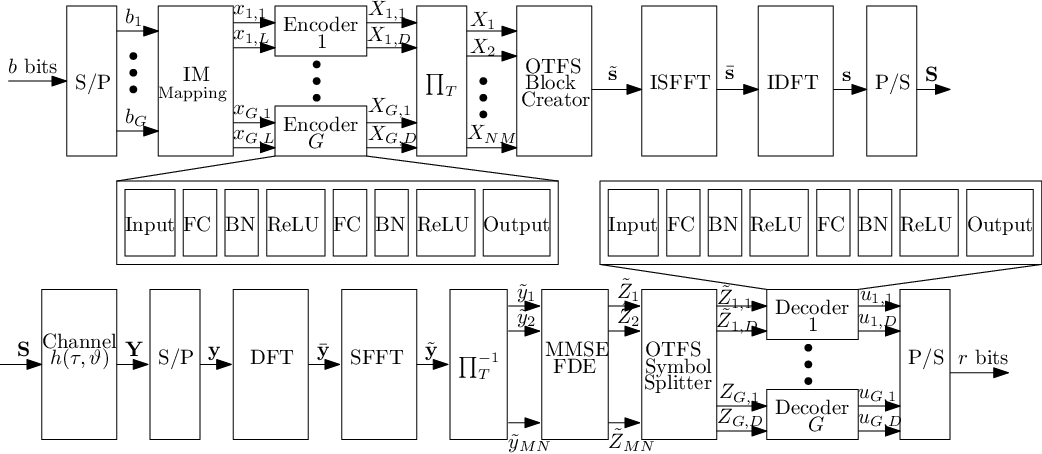}}%
\caption{Block diagram of the proposed AE-based {OTFS-IM} system.}
\label{fig:aedotfs}
\end{figure*}

\subsection{{PAPR} of MB-DFT-S-OTFS-IM Transmit Signal}
 For conventional {OTFS} signal, the {PAPR} can be represented as:
\begin{align}\label{paprtime}
\text{PAPR} = \frac{\max\left[|s[n,m]|^2\right]}{\mathbb{E}\left[|s[n,m]|^2\right]},
\end{align}
where $\mathbb{E}[\cdot]$ denotes the expectation operator. In the case of {OTFS} modulation, assuming an ideal rectangular pulse shaping filter. We have:
\begin{align}\label{paprkldomain}
\text{PAPR} = \frac{N \cdot \max\left[|s[k,l]|^2\right]}{\mathbb{E}\left[|s[k,l]|^2\right]},
\end{align}
where $N$ is the number of delay bins in the {DD} domain. It can be observed from \eqref{paprkldomain} that the maximum PAPR of the {OTFS} transmit signal increases linearly with $N$, rather than $M$, which differs from traditional multi-carrier modulation techniques such as {OFDM}, where the {PAPR} typically scales with the number of subcarriers $M$.\\
Let $\mathbf{s} \in \mathbb{C}^{JMN \times 1}$ denote the $J$-times oversampled TD signal vector generated from the DD grid, where $J \ge 4$ is the oversampling factor required to capture the true continuous-time peaks. The {PAPR} of the transmitted {OTFS-IM} frame is fundamentally defined as:
\begin{align}
    \text{PAPR}(\mathbf{s}) \triangleq \frac{\max_{0 \le n \le JMN-1} |s[n]|^2}{\mathbb{E} \left[ |s[n]|^2 \right]},
    \label{eq:papr_def}
\end{align}
where $s[n]$ is the $n$-th element of $\mathbf{s}$, and $\mathbb{E}[\cdot]$ denotes the expected value, representing the average transmit power.
The absolute worst-case peak power of {OTFS-IM} is structurally bounded by the active ratio $\rho$:
\begin{align}
    \max \left( |s_{\text{IM}}[n]|^2 \right) \le \rho^2 \cdot \max \left( |s_{\text{std}}[n]|^2 \right),
    \label{eq:papr_bound}
\end{align}
where $s_{\text{std}}[n]$ represents the equivalent standard full-grid {OTFS} signal. 

The {MB-DFT-S-OTFS-IM} can be extend from (\ref{eq:dfts}) and (\ref{paprkldomain}) with the aid of DFT-spreading pre-coding as
\begin{align}
\text{PAPR} = \frac{ \max\sum^{G-1}_{g=1}\left[|\tilde x^{DD}[k,g]|^2\right]}{\mathbb{E}\left[|\tilde x^{DD}[k,g]|^2\right]}.
\end{align}
High {PAPR} may significantly distort the {OTFS} signal due to nonlinearities in the high-power amplifier, which in turn degrades the {BER} performance. Therefore, we aim for reducing the {PAPR} of the proposed system to improve {EE}, while maintaining satisfactory {BER} performance.
\section{PROPOSED AUTOENCODER}
We propose to reduce the {PAPR} by leveraging a combination of {AE} techniques and {DFT-S}. The  {AE} conceived is designed for learning an encoder function that maps the input {MB-DFT-S-OTFS-IM} signal to a low-{PAPR} representation, and a decoder function that reconstructs the original signal from this representation, while maintaining minimal {BER} degradation. We first describe the architecture of the proposed {AE}-based system, followed by the formulation of the loss function and the training procedure.

\subsection{Network Structure}
The block diagram of the proposed AE-based system is shown in Fig.~\ref{fig:aedotfs}. In the {MB-AE} model conceived, the {DD} domain {OTFS-IM} signal $\tilde x_g^{DD}$  of a single band is first encoded by the $g$-th encoder and the resultant output is expressed as $f_g(\mathbf{\tilde x^{DD}};\boldsymbol{\theta}^g_f)$, where $\boldsymbol{\theta}_f$ is the model coefficient of the encoder. Then the encoded symbols are interleaved and converted to the {TD} signal $\boldsymbol{s}$ by the of {ISFFT} and {IDFT} operator. The signal is then transmitted over the wireless channel, where the demodulator converts the received signal $\boldsymbol{y}$ into the {DD} signal $ \boldsymbol{\tilde y}$. Following channel equalization, the equalized signals after de-interleaving are decoded by $G$ decoders. Then the reconstructed symbol can be denoted by $g_g(\mathbf{\tilde z};\boldsymbol{\theta}^g_g)$, where $\boldsymbol{\theta}^g_g$ is the model coefficient of the $g$-th decoder.
From a practical perspective, the MB-AE offers several advantages over its single-band counterpart. By operating on lower-dimensional sub-band signals, the MB-AE significantly reduces the sub-network input size, which overcome the gradient vanishing issues commonly observed in large single-band autoencoders. Then, the number of sub-bands of MB-AE can be adjusted to trade off performance, complexity, and latency.
Specifically, the {TD} signal and the reconstructed output can be expressed as
\begin{align}
\mathbf{s} &= \mathrm{IFFT}\!(\mathrm{ISFFT}(\Pi(\sum^{G}_{g=1} f_g(\mathbf{\tilde x^{DD}};\boldsymbol{\theta}^g_f))), \label{eq:sig_enc}\\ 
\hat{\mathbf{x}} &= \sum^{G}_{g=1} g_g\!\left(\mathrm{SFFT}\!\left(\mathrm{FFT}(\Pi^{-1}(\mathbf{Z})\right);\boldsymbol{\theta}^g_g\right).\label{eq:sig_dec}
\end{align}

Accordingly, the symbol $u$ reconstructed at the receiver, can be written in a stylized format as follows:
\begin{align}
\boldsymbol{u}=g\circ\Pi^{-1}\circ \text{SFFT}\circ \text{DFT} \circ \mathbb{H}\circ \text{ISFFT}\circ \text{IDFT}\circ \Pi\circ f(\textbf{x}),
\end{align}
where $\mathbb{H}$ represents the effect of the fading channel and {AWGN}, while $\Pi$ is the interleaver function.
Combining the {OTFS} input–output equivalent relationship (\ref{eq:oimatrix}), the AE model can be written compactly as
\begin{align}
\hat{\mathbf{x}} = g[H_\mathrm{e}f(\mathbf{x};\boldsymbol{\theta}_f) + \tilde{\mathbf{w}};\boldsymbol{\theta}_g].
\label{eq:ae_compact}
\end{align}
In the following, we will introduce the encoder and decoder, each of which consists of four layers: an input layer, two hidden layers, and an output layer. 
Each hidden layer contains a Fully Connected ({FC}) layer, a Rectified Linear Unit ({ReLU}) activation, and a Batch Normalization ({BN}) layer. 
For the encoder, the output of the $m$-th {FC} layer is
\begin{align}
\mathbf{y}^{m}_{\mathrm{FC}} = \boldsymbol{W}^{(f)}_m \mathbf{x}^{m}_{\mathrm{FC}} + \mathbf{b}^{(f)}_m,
\end{align}
where $\mathbf{x}^{m}$ is the input data, while $\boldsymbol{W}^{(f)}_m$ and $\mathbf{b}^{(f)}_m$ denote the weights and biases, respectively. 
Similarly, $\boldsymbol{W}^{(g)}_m$ and $\mathbf{b}^{(g)}_m$ denote the weights and bias for the {FC} layer of the decoder. The number of neurons in the {FC} layer is set to $3MN$ in this work, which includes the real and imagine part plus the power factor of the input data. Again, the {ReLU}, denoted by the $\sigma(\cdot)$ is used as activation function, followed by the {BN} layer normalizing the output of the {ReLU} function for preventing small changes in the encoder and decoder parameters causing suboptimal perturbations of the gradients.
The {BN} operation is formulated as:
\begin{align}
\mathrm{BN}(\sigma(\mathbf{y}^m_{\mathrm{FC}})) = 
\gamma \frac{\sigma(\mathbf{y}^m_{\mathrm{FC}}) - \mathbb{E}[\sigma(\mathbf{y}^m_{\mathrm{FC}})]}
{\sqrt{\mathrm{Var}[\sigma(\mathbf{y}^m_{\mathrm{FC}})] + \nu}} + \beta,
\end{align}
where $\gamma$ and $\beta$ are the scaling and shifting factors, $\nu$ is a small constant to prevent division by zero, and $\sigma(\cdot)$ denotes the {ReLU} activation.
\vspace{-3mm}
\subsection{Loss Function Design}
Again, the {AE} is trained to reduce {PAPR} while maintaining a low {BER}.  Thus, a pair of objectives are considered: (i) reconstructing the original signal $\mathbf{x}$, and  (ii) generating a low-{PAPR} waveform $\mathbf{s}$.

The reconstruction loss minimizing the {BER} is formulated as:

\begin{align}
\mathcal{L}_1 = \|\mathbf{x} - g(H_\mathrm{e} f(\mathbf{x};\boldsymbol{\theta}_f) + \tilde{\mathbf{w}};\boldsymbol{\theta}_g)\|_2^2, \label{eq:loss1}
\end{align}
While the {PAPR} loss is defined as

\begin{align}
\mathcal{L}_2 = \mathrm{PAPR}(\mathbf{s}) 
= \frac{\|\mathbf{s}\|_\infty^2}{\mathbb{E}[\|\mathbf{s}\|_2^2]/(MN)}. \label{eq:loss2}
\end{align}
The total loss function is a weighted sum:

\begin{align}
\mathcal{L} = \mathcal{L}_1 + \eta \mathcal{L}_2, \label{eq:total_loss_ae}
\end{align}
where $\eta$ is a hyperparameter controlling the trade-off between erosion {BER} and {PAPR} reduction. 
A lower $\eta$ improves {BER} at the expense of higher {PAPR}, while a higher $\eta$ provides stronger {PAPR} suppression.

\subsection{Training Process}  

A two-step training procedure is adopted for ensuring convergence stability.

\begin{itemize}
    \item \textbf{Step 1 (Pretraining):} Initialize the network using a uniformly distributed random initializer and train it by minimizing $\mathcal{L}_1$ only. This allows the {AE} to accurately reconstruct $\mathbf{x}$ before focusing on {PAPR} reduction.
    \item \textbf{Step 2 (Twin-task training):} Using the pretrained parameters for initialization, train the {AE} by minimizing the combined loss $\mathcal{L}$ in~\eqref{eq:total_loss_ae}.
\end{itemize}

During training, the parameters are updated via the Adam optimizer as

\begin{align}
\boldsymbol{\theta}^{(k+1)} = \boldsymbol{\theta}^{(k)} - \mu \nabla_{\boldsymbol{\theta}} \mathcal{L},
\end{align}
where $\mu$ is the learning rate. 
Training proceeds until the maximum affordable number of iterations $K$ is reached or the loss function converges.

\begin{algorithm}[!t]
\caption{Training Procedure for {AE}-Based {PAPR} Reduction}
\label{alg:ae_training}
\begin{algorithmic}[1]
\REQUIRE Maximum iterations $K_1$, $K_2$, learning rate $\mu$, hyperparameter $\eta$, training {SNR}.
\ENSURE Trained parameters $\boldsymbol{\theta}_f^\star$, $\boldsymbol{\theta}_g^\star$.
\STATE Initialize $\boldsymbol{\theta}^{(0)}$ using uniform distribution.
\FOR{$k_1=1$ to $K_1$}
    \STATE Compute $\mathcal{L}_1$ from~\eqref{eq:loss1}.
    \STATE Update $\boldsymbol{\theta}^{(k)} \leftarrow \boldsymbol{\theta}^{(k-1)} - \mu\nabla\mathcal{L}_1$.
\ENDFOR
\FOR{$k_2=1$ to $K_2$}
    \STATE Compute total loss $\mathcal{L}$ from~\eqref{eq:total_loss_ae}.
    \STATE Update $\boldsymbol{\theta}^{(k)} \leftarrow \boldsymbol{\theta}^{(k-1)} - \mu\nabla\mathcal{L}$.
\ENDFOR
\STATE Output trained AE parameters $\boldsymbol{\theta}^\star = (\boldsymbol{\theta}_f^\star,\boldsymbol{\theta}_g^\star)$.
\end{algorithmic}
\end{algorithm}

\section{SD DETECTION FOR CHANNEL DECODING}
{SD} detection aims for achieving near-capacity performance when combined with advanced channel coding schemes such as Low-Density Parity-Check ({LDPC}) codes, either with or without iterative soft-information exchange between the {MIMO} demapper and the channel decoder.
The complexity of the optimal Maximum \textit{a posteriori} (MAP) MIMO detector becomes prohibitive as the modulation order and number of transmit antennas increase \cite{Nguyen12}. 
Therefore, typically suboptimal Soft-Input Soft-Output (SISO) detectors are employed, such as soft demodulators based on the Approximate Log-MAP algorithm or message passing methods\cite{jmim}.
In light of these considerations, we conceive learning-aided {SD} detection for the proposed {OTFS-IM} system to approach near-optimal performance at a moderate complexity.
\begin{figure}[!htb]
\centering

{\includegraphics[width=7.8cm]{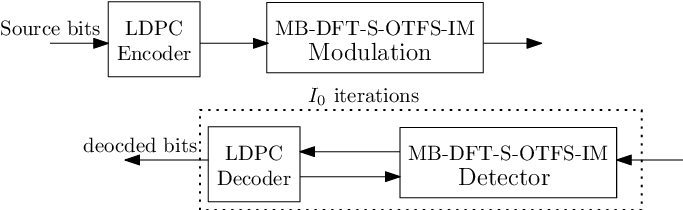}}%
\caption{The transceiver architecture of LDPC encoded MB-DFT-S-OTFS-IM}
\label{fig:ldpc}

\end{figure}

\subsection{LDPC-Coded SD Detection}
In the proposed LDPC-coded {OTFS-IM} receiver, the {SD} module computes bit-wise LLRs, which serve as soft inputs to the {LDPC} decoder.
The {LLR} for a coded bit $b$ is defined as

\begin{align}\label{eq:llr_def}
L(b) = \log\frac{p(b=1|\boldsymbol{Y})}{p(b=0|\boldsymbol{Y})}.
\end{align}

The {LLR} is defined as the probability ratio of the bit being '1' and '0', which can be written as $L(b)=log\frac{p(b=1)}{p(b=0)}$.
The conditional probability $p(\boldsymbol{Y}{|}\mathcal{X}_{\gamma,\beta,\varphi})$ of  receiving  the  group  signal $\textbf{Y}$ is given by \cite{bishop2006pattern}
\begin{equation}
\begin{aligned}
&p(\boldsymbol{Y}{|}\mathcal{X}_{\gamma,\beta})\\
&=\frac{1}{(\pi\boldsymbol{N}_0)^{NT}}exp(-\frac{||\boldsymbol{Y}-\boldsymbol{H}\boldsymbol{I}_{SI}(\gamma)\mathcal{X}_{q,l}(\beta)||^2}{N_0}).
\end{aligned}
\label{eq:cond}
\end{equation}

\begin{figure*}[t] %h
{\small
	\begin{equation}
	\begin{aligned}
\boldsymbol{L}_e(u_l)&=\ln\frac{\sum_{\mathcal{X}_{\gamma,\beta\in\mathcal{X}_1^l}}p(\boldsymbol{Y}|\mathcal{X}_{\gamma,\beta})exp[\sum_{j\neq{l}}u_jL_a(u_j)]}{\sum_{\mathcal{X}_{\gamma,\beta\in\mathcal{X}_0^l}}p(\boldsymbol{Y}|\mathcal{X}_{\gamma,\beta})exp[\sum_{j\neq{l}}b_jL_a(b_j)]}\\
&={\ln}\frac{\sum_{{\mathcal{X}_{{\gamma,\beta\in\mathcal{X}_1^l}}}}exp[-||\boldsymbol{Y}-\boldsymbol{H}\boldsymbol{\mathcal{I}}_{SI}(\gamma)\mathcal{X}_{q,l}(\beta)\|^2/N_0+\sum_{j\neq{l}}u_jL_a(u_j)]}{\sum_{\mathcal{X}_{\gamma,\beta\in\mathcal{X}_1^l}}exp[-\|\boldsymbol{Y}-\boldsymbol{H}\boldsymbol{\mathcal{I}}_{SI}(\gamma)\mathcal{X}_{q,l}(\beta)||^2/N_0+\sum_{j\neq{l}}u_jL_a(u_j)]}.
\end{aligned}
\label{eq:flst}
	\end{equation}}
		\hrulefill  %
\end{figure*}

Furthermore, $N_0$ is  the  noise  power, where  we  have $\sigma^{2}_n=N_0/2$. The equivalent received signal $\boldsymbol{Y}$ per subcarrier  group  carries $\boldsymbol{B}$ channel-coded bits $\boldsymbol{u}=[u_1,u_2,...,u_B]$ and the extrinsic {LLR} of bits $u_l ,(l=1,2,\cdots,B)$ is expressed by (\ref{eq:flst}) .
In (\ref{eq:flst}), ${{\mathcal{X}_1^l}}$ and ${\mathcal{X}_0^l} $ represent a subset of the legitimate  equivalent signal $\mathcal{X}$ corresponding to bit $u_l$ when $u_l=1$ and $u_l=0$, respectively, yielding $X_{1}^l \equiv \{{\mathcal{X}_{\gamma,\beta}\in\mathcal{X}:u_l=1}\}$ and $X_{0}^l \equiv \{\mathcal{X}_{\gamma,\beta}\in\mathcal{X}:u_l=0\}$. The variable ${L_a}(\cdot)$ in (\ref{eq:flst}) represents the \textit{a priori} {LLR} feedback from the {LDPC} decoder used. As shown in Fig.~\ref{fig:ldpc}, the {LLR} is exchange between the {LDPC} decoder and detector employing $I_0$ iterations.
\subsection{SD Based AE}
In this part, we leverage the {SD-AE} architecture conceived, in which the transmitter’s encoding network and the receiver’s decoding network are jointly trained to generate and exploit \emph{soft information  (LLRs)} rather than merely performing 'single-shot' hard symbol decisions. Specifically, instead of the conventional AE that simply maps input symbols to output estimates, our SD-AE is designed for ensuring that the decoder network outputs soft bit-wise likelihoods \(L(b)\) (cf. Eq.~\eqref{eq:llr_def}) for each coded bit. These {LLRs} are then fed directly into the {LDPC} channel decoder for iterative decoding. By using the {AE} to learn the mapping  
\[
\hat{\mathbf{u}} = \mathrm{Decoder}\bigl( \mathbf{y}; \boldsymbol{\theta}_g \bigr),
\]
where \(\mathbf{y}\) is the received signal and the decoder produces soft bit outputs \(L(b)\), the network learns to approximate the \textit{a posteriori} probabilities  \vspace{-2mm}
\[
p(b = 1 \mid \mathbf{y}) \quad \text{and} \quad p(b = 0 \mid \mathbf{y}),
\]
and thus generates soft inputs for the channel decoder using the Softmax layer as its output. The SD-AE aim for producing {LLR}s that maximize the iterative soft decoding performance.\\
To intrinsically integrate our model with the subsequent {LDPC} channel decoder, the network's decoding module is fundamentally modified. Specifically, the final linear output layer of the {HD-AE} is replaced with a Softmax layer. For a localized sub-band consisting of $N_f$ active grids utilizing an $\mathcal{M}$-ary constellation, the output dimension of this final dense layer is explicitly expanded to $N_f \times \mathcal{M}$. Consequently, instead of outputting continuous values mapped to $(I, Q)$ coordinates, the Softmax activation ensures the network directly outputs a discrete \textit{a posteriori} probability distribution for each active grid, which can be convert to LLRs. Crucially, to facilitate this probabilistic learning, the loss function must be adapted. The MSE reconstruction loss is replaced entirely with the cross entropy loss. 

\vspace{-2mm}
\begin{table*}[t]
\centering

\caption{Architecture of the Proposed DNN Encoder–OFDM–Decoder System}
\label{tab:DNN_structure}
\scalebox{0.88}{
\begin{tabular}{c|c|c|c|c}
\hline
\textbf{Module} & \textbf{Layer} & \textbf{Input Size} & \textbf{Output Size} & \textbf{Operation / Activation} \\
\hline
{Encoder}
& Input            & $MN/H$ complex ($2MN/G$ real) & $2MN/G$ & Real/imag. concatenation \\
& FC-1             & $2MN/G$ & $Q^{\text{enc}}_1$ & Linear + BatchNorm + ReLU \\
& FC-2             & $Q^{\text{enc}}_1$ & $Q^{\text{enc}}_2$  & Linear \\
& FC-3 (Output)   & $Q^{\text{enc}}_2$& $MN/G$  complex & Power normalization + Complex reconstruction \\
\hline
{OTFS} In-and-Out
& Interleaver             &$G*2MN/G$ complex & $MN$ complex & interleave encoders \\
& Channel          & $MN$ complex & $MN$ complex & Doubly channel/{NTN} channel fading + noise \\
& De-Interleaver              & $MN$ complex & $G*2MN/G$ complex & De-interleave decoders \\
\hline
{Decoder}
& Input            & $2MN/G$ complex + $MN/G$$|H_n|^2$ & $3MN/G$ real & Feature construction \\
& FC-1             & $3MN/G$ & $Q^{\text{dec}}_1$ & Linear + BatchNorm + ReLU \\
& FC-2             & $Q^{\text{dec}}_1$ & $Q^{\text{dec}}_2$ & Linear + BatchNorm + ReLU \\
& FC-3             & $Q^{\text{dec}}_2$ & $Q^{\text{dec}}_3$ & Linear + BatchNorm + ReLU \\
& FC-4(Output)       & $Q^{\text{dec}}_3$ & $2MN/G$ & Linear, complex reconstruction \\
\hline
\end{tabular}
}
\vspace{-2mm}
\end{table*}

\begin{table}[!htb]
\vspace{-2mm}
\centering
\caption{Simulation Parameters}
\label{tab:sim_param}
\scalebox{0.9}{
\begin{tabular}{l c}
\hline\hline
\textbf{Parameter}                & \textbf{Value} \\
\hline
Carrier frequency                 & 25.675 GHz \\
{SCS}                & 90 kHz \\
FIM Group size                     & 16,64,128 \\
FIM inactive IM subcarrier                   & 2 \\
GFIM Group size                     & 8 \\
GFIM Group number                    & 2,8,16 \\
Modulation scheme                 & (4,16)-QAM \\
{DD} grid subcarrier size M      & 16,64,128 \\
{DD} grid Time slot size N    & 16 \\
Number of channel paths \(P\)     & 10 \\
Path power coefficients \(\rho_i\)& \(\{\rho_i\}_{i=0}^{P-1}\) \\
Path Doppler shift \(\nu_i\)      & \(\nu_i = \nu_{\max}\cos \alpha_i\) \\
Satellite relative altitude                         & 300 km \\
Satellite relative speed                          & 7433 m/s \\
\hline\hline
LDPC code length                  & 8192 \\
LDPC code rate                    & \(1/3\) \\
LDPC decoding algorithm           & Sum-Product Algorithm \\
\hline\hline
\end{tabular}
}

\end{table}
\vspace{-1mm}
\section{SIMULATION RESULTS AND ANALYSIS}
\begin{figure}[!htb]
    \centering
    \vspace{-1mm}
    \includegraphics[width=0.48\textwidth]{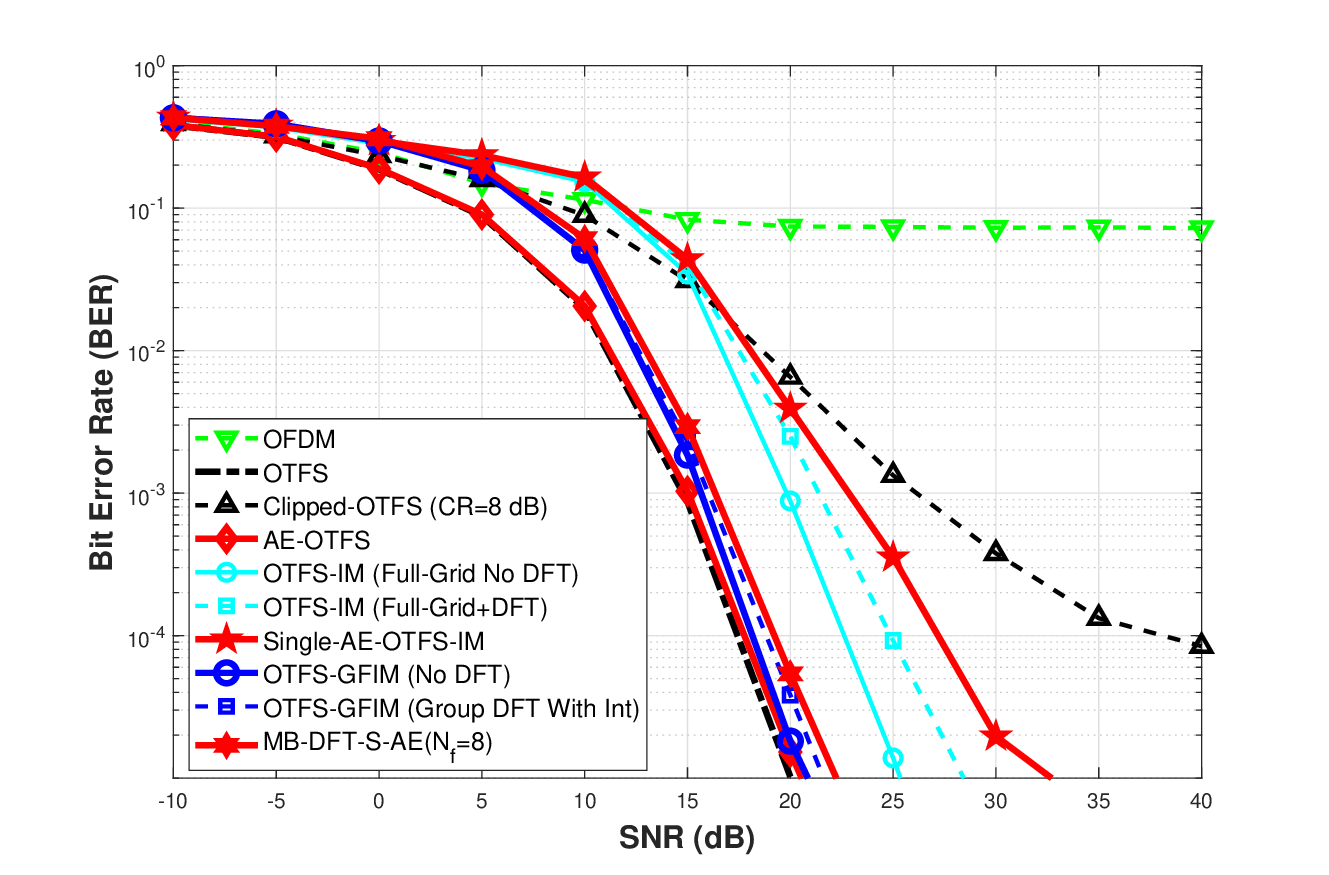}
    \caption{BER performance comparison between the proposed MB-AE-based scheme and traditional benchmark detectors over a shadowed-Rician NTN channel with $M=N=16$, $K_{null}=2$,$\mathcal{L}= 4QAM$, $G=2$}
    \vspace{-3mm}
    \label{fig:benchber}

\end{figure}
\begin{figure}[!htb]
    \centering
    \vspace{-4mm}
    \includegraphics[width=0.48\textwidth]{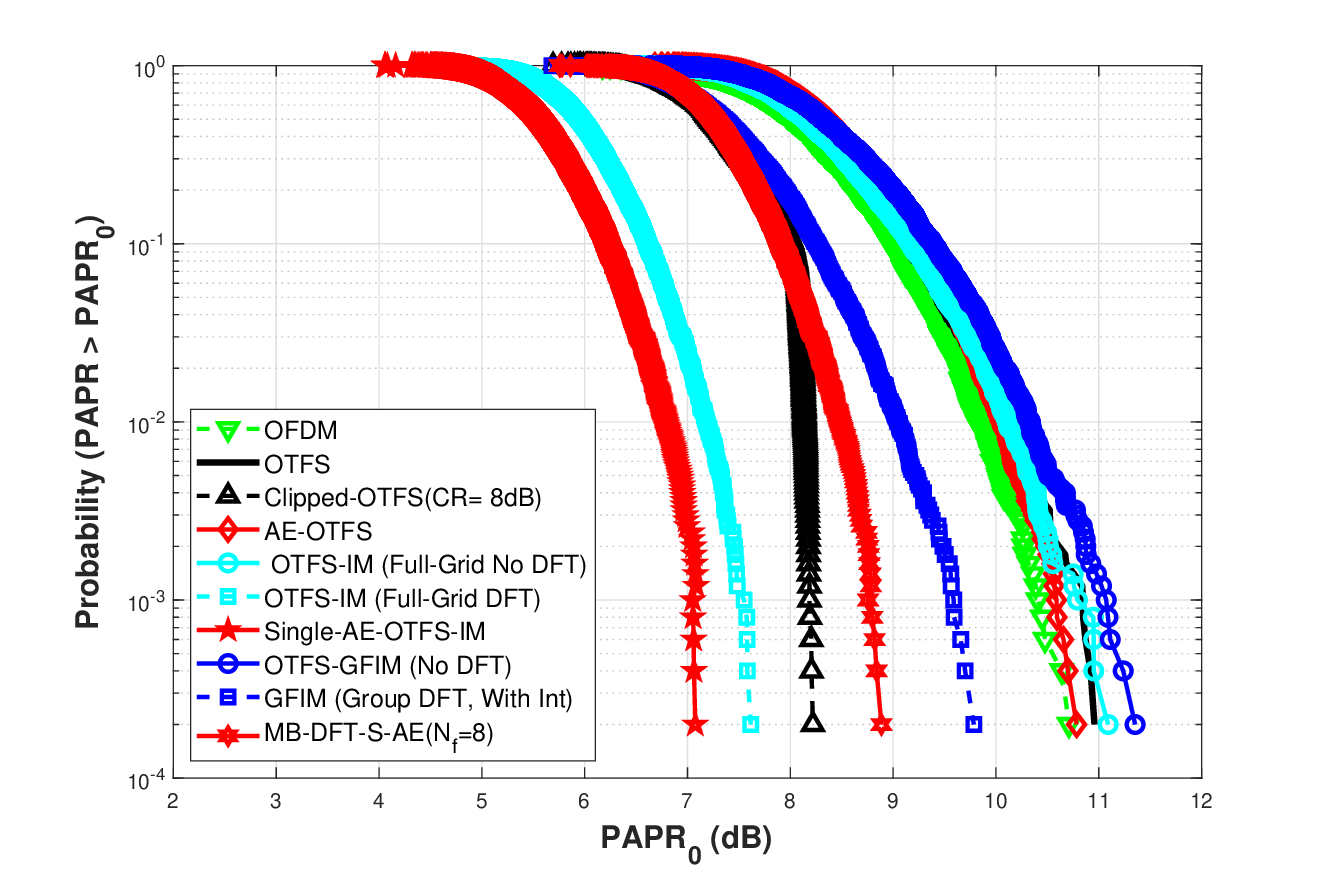}
    \caption{PAPR comparison between the proposed MB-AE-based scheme and traditional benchmark detectors over a shadowed-Rician NTN channel with $M=N=16$, $K_{null}=2$,$\mathcal{L}= 4QAM$, $G=2$}
    \vspace{-3mm}
    \label{fig:benchpapr}

\end{figure}
In this section, we evaluate the {PAPR} and {BER} performance of the proposed {AE}-based {PAPR} reduction scheme through numerical simulations. The simulation parameters are as follows: the carrier frequency is set to 25.675~GHz, the {SCS} is 90~kHz, and 4-Quadrature Amplitude Modulation ({QAM}) is employed. An {OTFS-IM} scheme is considered with a {DD} grid of size \( M = 16 \) and \( N = 16 \). 

The channel has \( P = 10 \) multi-path components with an exponentially decaying power delay profile and Jakes’ Doppler spectrum. The power delay profile is denoted as \( \{ \rho_i \}_{i=0}^{P-1} \), and the Doppler shifts \( \nu_i \) associated with each path are generated according to\vspace{-2mm}
\[\vspace{-1mm}
\nu_i = \nu_{\text{max}} \cos \alpha_i,
\]
where \( \nu_{\text{max}} \) represents the maximum Doppler shift determined by the user equipment speed. The angles \( \alpha_i \) are drawn uniformly from the interval \( [-\pi, \pi] \). A high-mobility scenario is simulated by assuming a LEO satellite at an altitude of 300~km. According to the relative speed calculation in \cite{ntnchannel}, the corresponding relative velocity between the satellite and the ground terminal is calculated to be approximately 7433~m/s.

The proposed autoencoder-based {PAPR} reduction module builds upon the traditional {AE} framework  paired with the GFIM technique. By leveraging this MB architecture, the inherently prohibitive complexity of applying a single, massive {AE} across the entire {DD} grid is circumvented. Instead, the highly complex global {AE} is effectively decomposed into a set of $G$ generalized, lightweight, and parallel small {AE}s. Specifically, an input vector $\textbf{s}\in\mathbb{C}^{MN\times 1}$ is split into $G$ sub-groups, each processed by a distinct sub-encoder at the transmitter and a corresponding sub-decoder at the receiver. For each sub-encoder, the input is converted into a $2MN/G$-dimensional real representation, followed by two Fully Connected ({FC}) layers harnessing both Batch Normalization ({BN}) and {ReLU} activation. The final encoder layer outputs a power-normalized complex-valued vector for each sub-group.

Then, in the {OTFS} module of Fig. \ref{fig:aedotfs}, the outputs of the multiple parallel encoders are interleaved to form a composite {OTFS} symbol of size $MN$. This interleaving operation ensures that each encoder contributes to a distributed part of the {DD} grid, while maintaining the global {OTFS} structure. The resultant {OTFS} symbol is then passed through the doubly selective {NTN} channel, after which a de-interleaver decomposes the received {OTFS} frame back into $G$ branches for processing by the corresponding decoders.

At the receiver side, each decoder of Fig. \ref{fig:aedotfs} operates on its corresponding received {OTFS} sub-block, constructing a $3MN/G$-dimensional real feature vector. Each decoder consists of three {FC} layers with {BN} and {ReLU} activation, followed by a linear output layer that reconstructs the real and imaginary components of the transmitted symbols.  Again, the neural networks are trained jointly using 80,000 samples, with a batch size of 200, and using the Adam optimizer with a learning rate of 0.001. The proposed design strikes an optimal balance between training stability, model expressiveness, and a lightweight per-block computational structure. The detailed configuration of the encoder–decoder network is summarized in Table~\ref{tab:DNN_structure}.

Conventional OTFS modulation inherently suffers from high PAPR due to its 2D spreading nature in Fig.~\ref{fig:aedotfs}. To address this, the OTFS-IM architecture combined with DFT spreading is considered. To ensure fair comparisons at an equivalent effective throughput, OTFS-IM employs null subcarrier-based IM. While this introduces a slight BER penalty compared to standard OTFS in Fig.~\ref{fig:benchber}, it leverages index activation sparsity for implicitly reducing PAPR. Furthermore, introducing DFT spreading  to uniformly distribute signal energy, smoothing the TD envelope and yielding superior PAPR reduction as shown in Fig.\ref{fig:benchpapr}.

To evaluate the effective transmission rates, we formulate the SE for each scheme. In standard OTFS, each grid. Slots of the DD domain conveys an $\mathcal{M}$-ary constellation symbol, yielding a constant SE:
\begin{align}\label{eq:se_std}
    SE_{\text{std}} &= \log_2(\mathcal{M}).
\end{align}

The SE of IM-based architectures takes into account both the constellation and spatial index bits. For the full-grid OTFS-IM baseline, IM is applied globally across the $MN$ grid with a single silent grid ($K_{z,\text{full}}=1$), resulting in:
\begin{align}\label{eq:se_full}
    SE_{\text{full}} &= \frac{\lfloor \log_2 \binom{MN}{K_{z,\text{full}}} \rfloor + K_{a,\text{full}} \log_2(\mathcal{M})}{MN}.
\end{align}
With $MN=256$, the combinatorial space provides $\lfloor \log_2 \binom{256}{1} \rfloor = 8$ index bits. Added to the $255 \times 2 = 510$ conventional modulation bits (assuming $\mathcal{M}=4$), the effective SE is $518 / 256 \approx 2.02$~bits/s/Hz.

For the grouped OTFS-GFIM scheme, the DD grid is partitioned into subgroups of size $N_f$. Within each subgroup, $K_{z,\text{gfim}}$ grids are nulled, and the remaining $K_{a,\text{gfim}} = N_f - K_{z,\text{gfim}}$ grids carry $\mathcal{M}$-QAM symbols. The SE is formulated as:
\begin{align}\label{eq:se_gfim}
    SE_{\text{gfim}} &= \frac{\lfloor \log_2 \binom{N_f}{K_{z,\text{gfim}}} \rfloor + K_{a,\text{gfim}} \log_2(\mathcal{M})}{N_f}.
\end{align}
Substituting our parameters of $N_f=8$, $K_{z,\text{gfim}}=2$, $\mathcal{M}=4$, the subgroup yields $\lfloor \log_2 \binom{8}{2} \rfloor = 4$ index bits. Combined with $6 \times 2 = 12$ modulation bits, each subgroup carries 16 bits over 8 grids slots. Thus, the SE of OTFS-GFIM perfectly aligns with that of standard OTFS at exactly $2.0$~bits/s/Hz.

We first benchmark the proposed architecture against standard {OFDM}, {OTFS}, and Clipped {OTFS} in Fig.~\ref{fig:benchber} and \ref{fig:benchpapr}. Conventional {OTFS} exhibits a high {PAPR} comparable to that of {OFDM}. While hard clipping drastically lowers {PAPR}, it introduces severe non-linear distortion that degrades signal integrity, yielding an a substantial {BER} floor.
\begin{figure*}[!htb]
\vspace{-8mm}
\centering                                
\subfigure[BER performance $M=64,N=16$, $K_{null}=2$,$\mathcal{L}= 4QAM$, $G=8$.]{                    
\begin{minipage}{5.6cm}
\includegraphics[width=5.4cm]{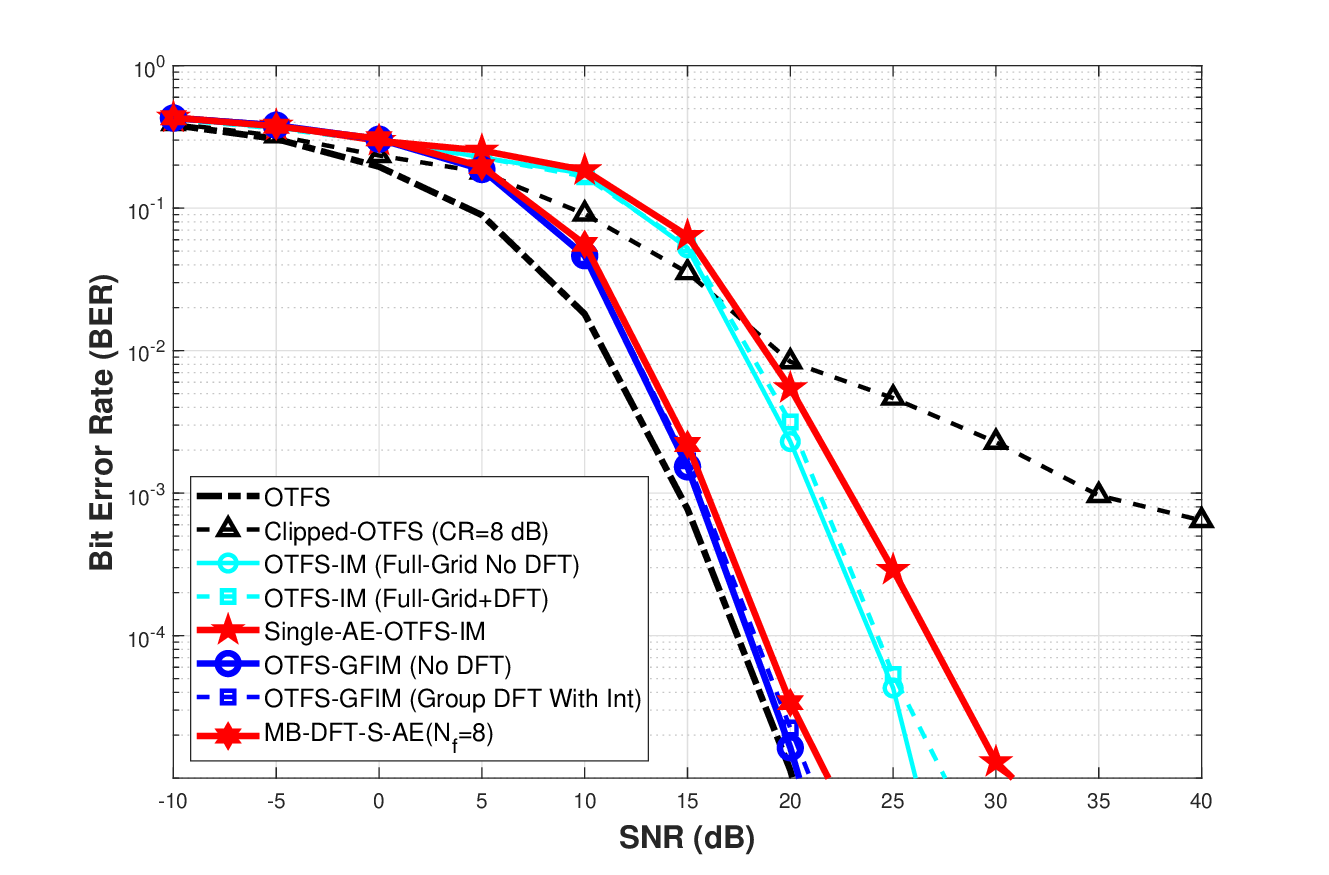}    
\label{fig:s11_ber} 
\end{minipage}}
\subfigure[BER performance $M=128,N=16$, $K_{null}=2$,$\mathcal{L}= 4QAM$, $G=16$.]{                    
\begin{minipage}{5.6cm}
\includegraphics[width=5.4cm]{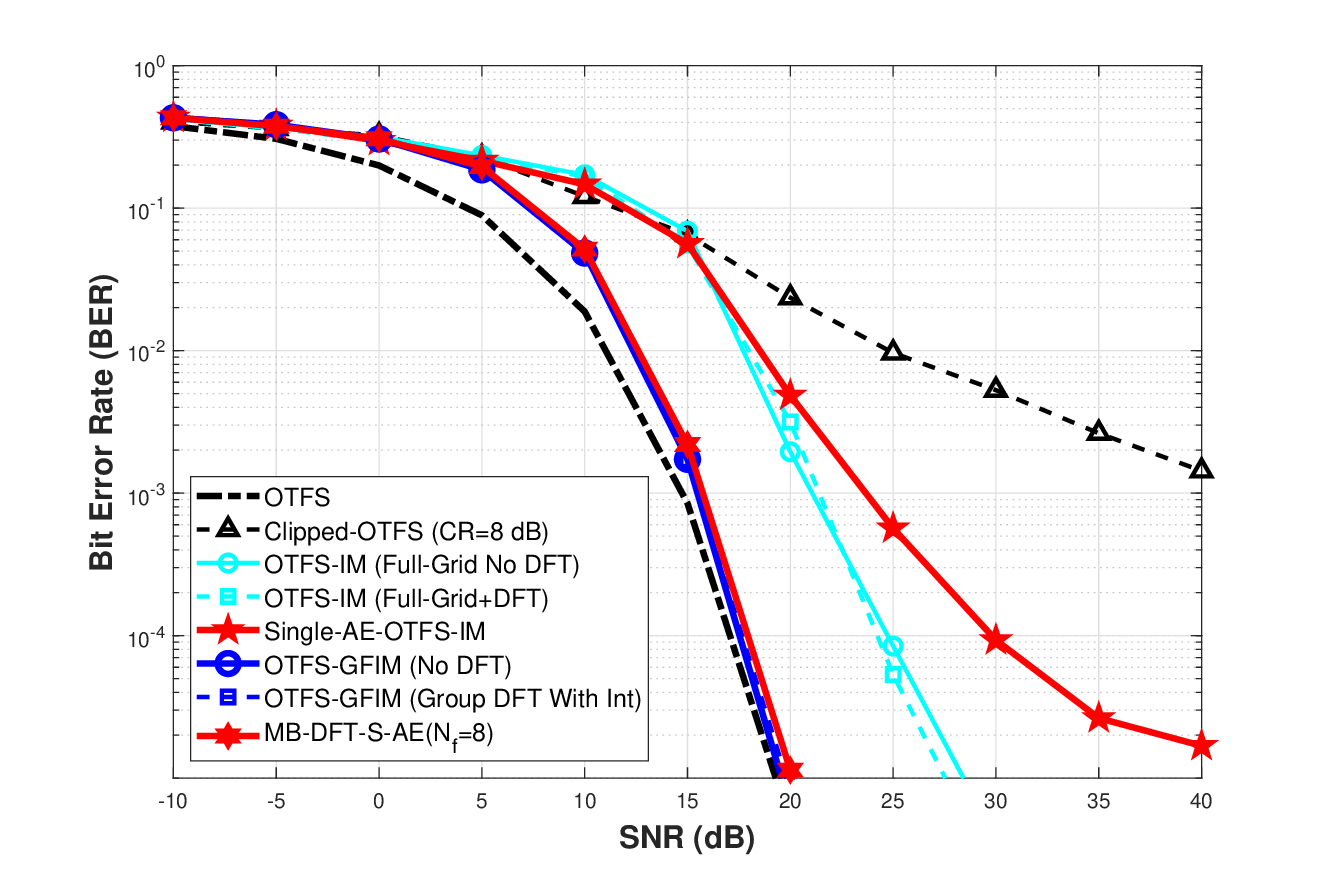}
\label{fig:s21_ber}
\end{minipage}}
\subfigure[ BER performance $M=16,N=16$, $K_{null}=2$,$\mathcal{L}= 16QAM$, $G=2$.]{ 
\begin{minipage}{5.6cm}
\includegraphics[width=5cm]{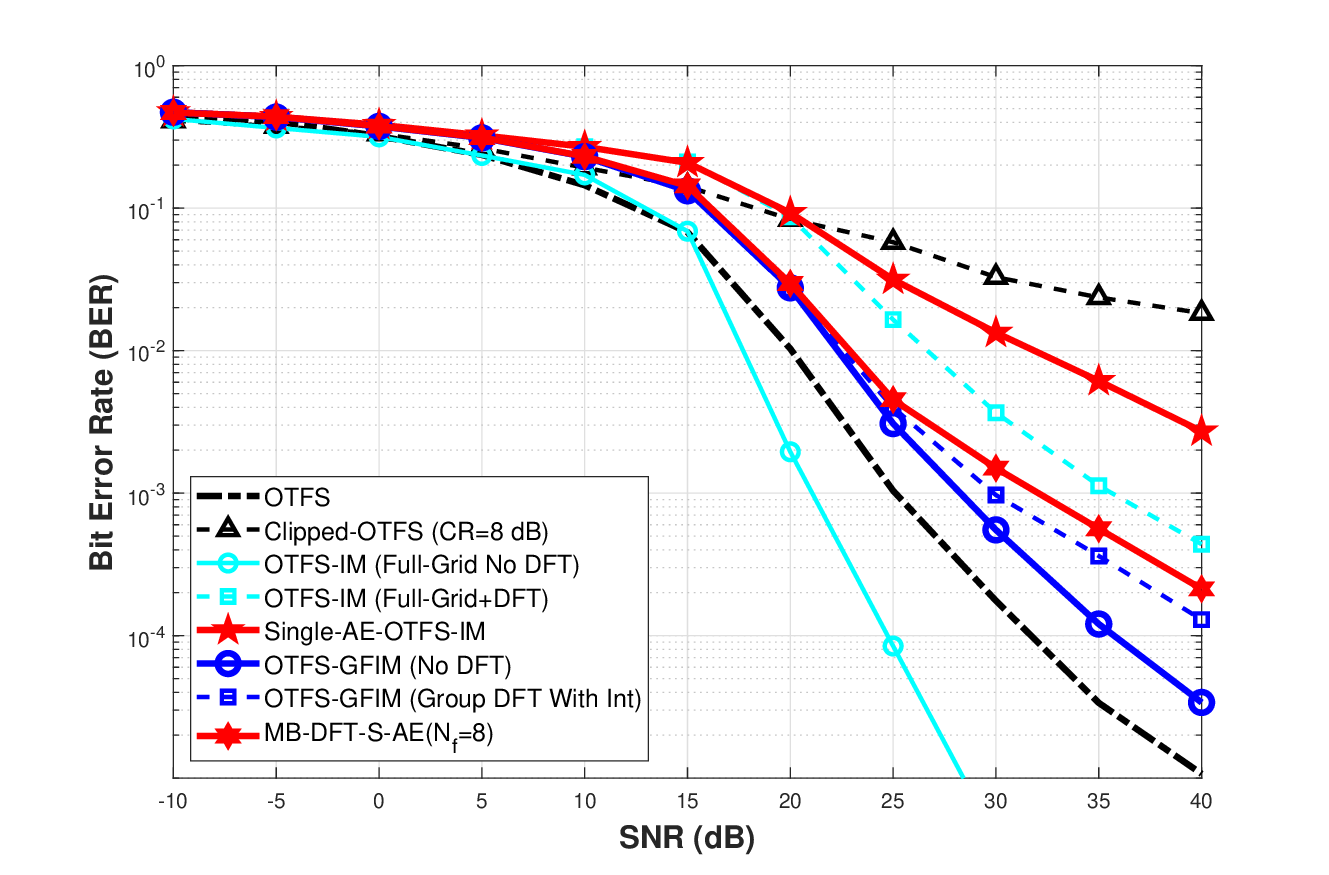}
\label{fig:s31_ber}
\end{minipage}}
\centering                          
\subfigure[CCDF of PAPR $M=64,N=16$, $K_{null}=2$,$\mathcal{L}= 4QAM$, $G=8$.]{                    
\begin{minipage}{5.6cm}
\includegraphics[width=5.4cm]{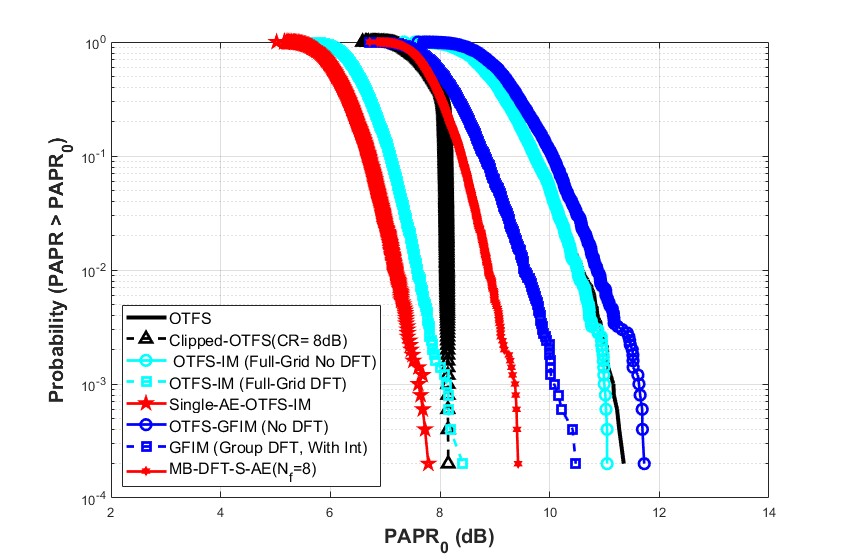}    
\label{fig:s12_papr} 
\end{minipage}}
\subfigure[CCDF of PAPR $M=128,N=16$, $K_{null}=2$,$\mathcal{L}= 4QAM$, $G=16$.]{                    
\begin{minipage}{5.6cm}
\includegraphics[width=5.4cm]{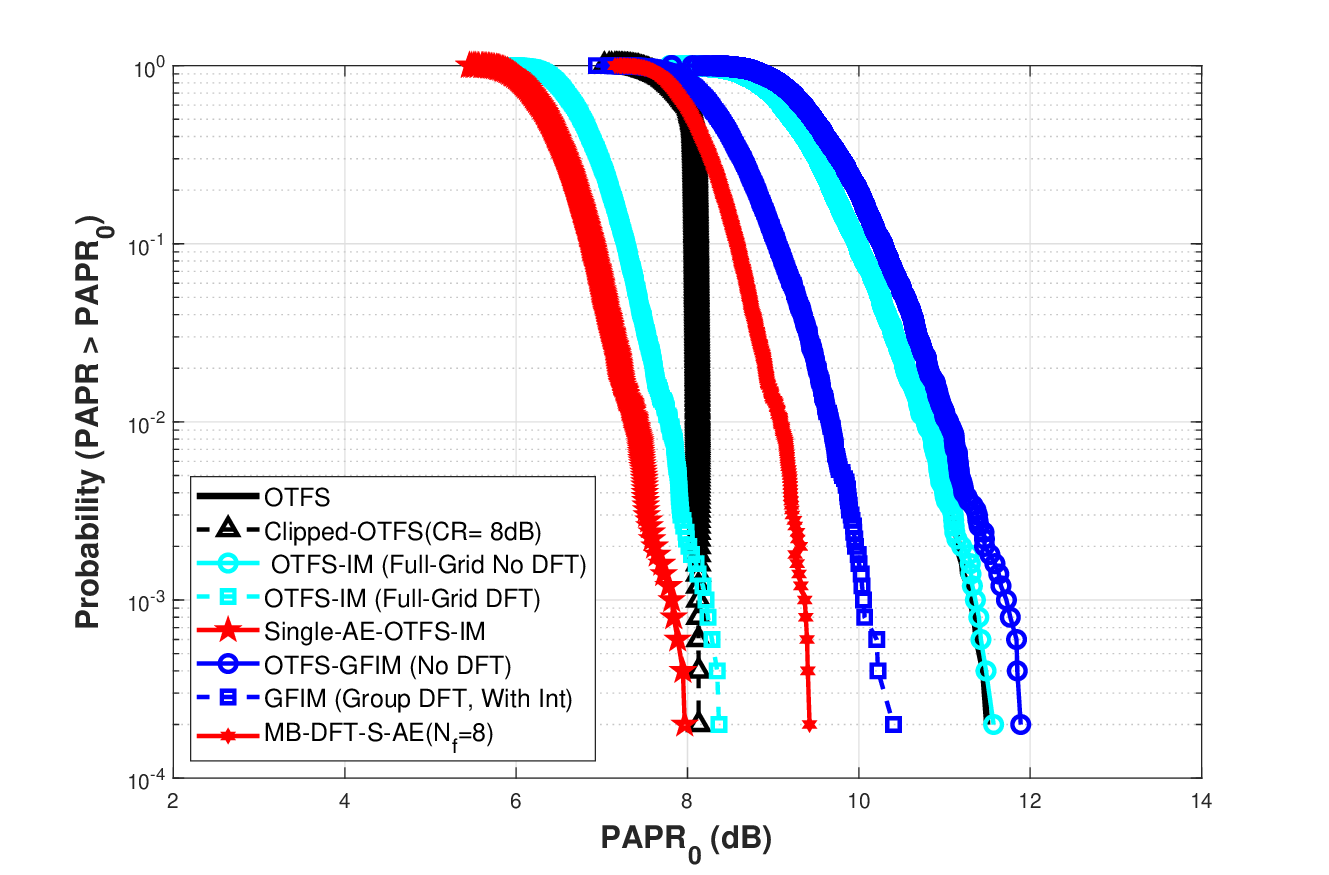}
\label{fig:s22_papr}
\end{minipage}}
\subfigure[CCDF of PAPR $M=16,N=16$, $K_{null}=2$,$\mathcal{L}= 16QAM$, $G=2$.]{ 
\begin{minipage}{5.6cm}
\includegraphics[width=5.4cm]{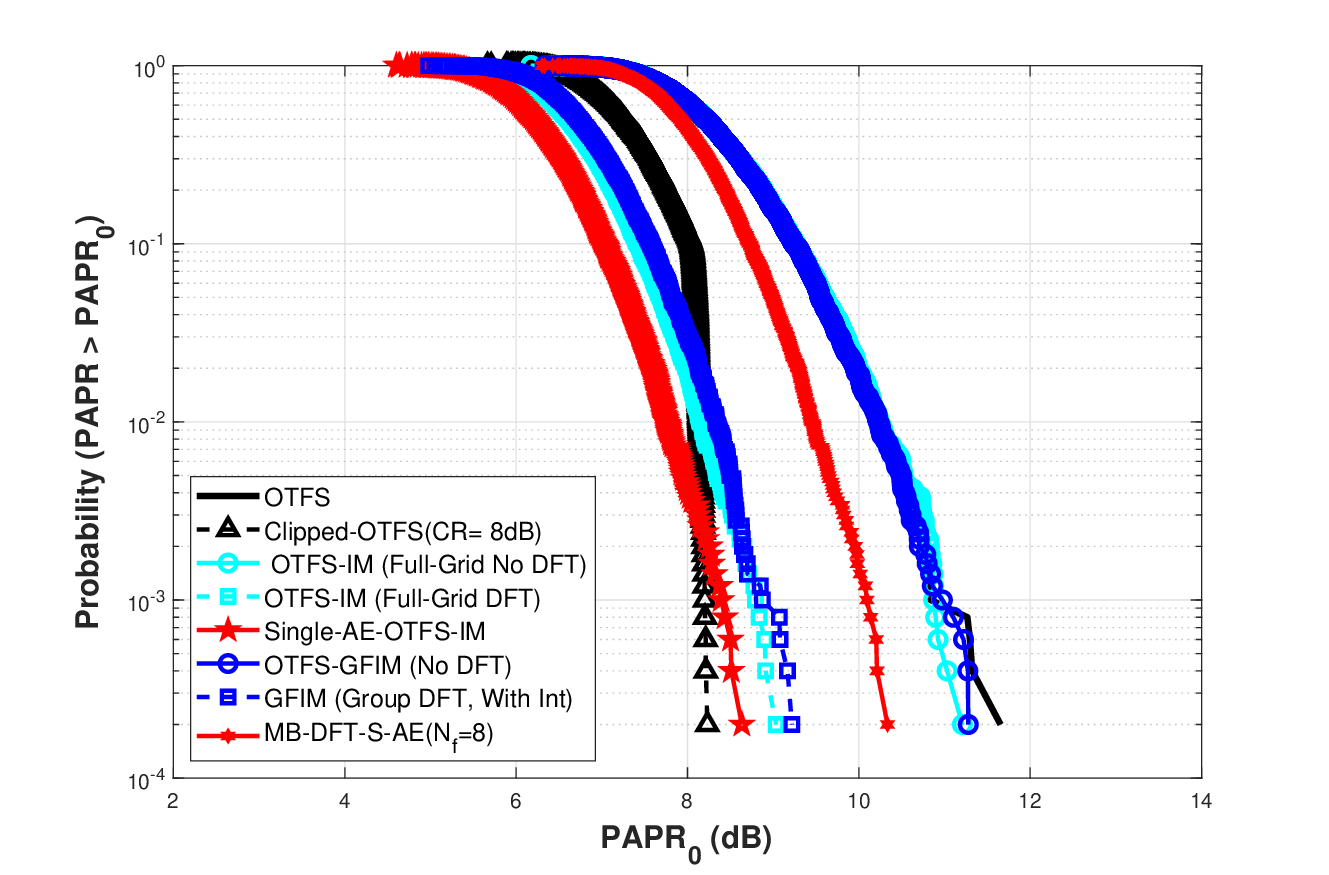}
\label{fig:s32_papr}
\end{minipage}}
\centering

\caption{Illustration of the BER performance and CCDF of PAPR comparison between the proposed MB-AE-based scheme and  benchmark over a shadowed-Rician NTN channel.}  
\vspace{-6mm}
\end{figure*}

Although the {AE} architecture lowers {PAPR} via non-linear constellation shaping while approaching the optimal {BER}, finding the ideal configuration requires a delicate trade-off. For this baseline comparison, the weighting hyperparameter is strategically fixed at $\eta=0.01$ to reduce {PAPR} without compromising the {BER}. Furthermore, a detailed $\eta$ sensitivity analysis follows in subsequent sections. Applying the {AE} directly to standard {OTFS} ({AE-OTFS}) maintains excellent {BER}, but its {PAPR} remains relatively high. Conversely, applying the {AE} to a full-grid {OTFS-IM} architecture with {DFT} spreading aggressively suppresses {PAPR}, but the drastic reduction in active subcarriers severely restricts the Euclidean distance, resulting in excessive {BER}.

Therefore, the proposed {MB-AE} emerges as the optimal solution. By partitioning the grid, {MB-AE} circumvents the performance penalties of a massive single grid, successfully maintaining robust {BER} performance, while achieving an exceptionally low {PAPR} that rivals Clipped {OTFS}.

\begin{figure*}[!htb]
\vspace{-7mm}
\centering                                
\subfigure[BER performance of MB-AE-DFT-S-OTFS in comparison to OTFS.]{                    
\begin{minipage}{5.6cm}
\includegraphics[width=5.4cm]{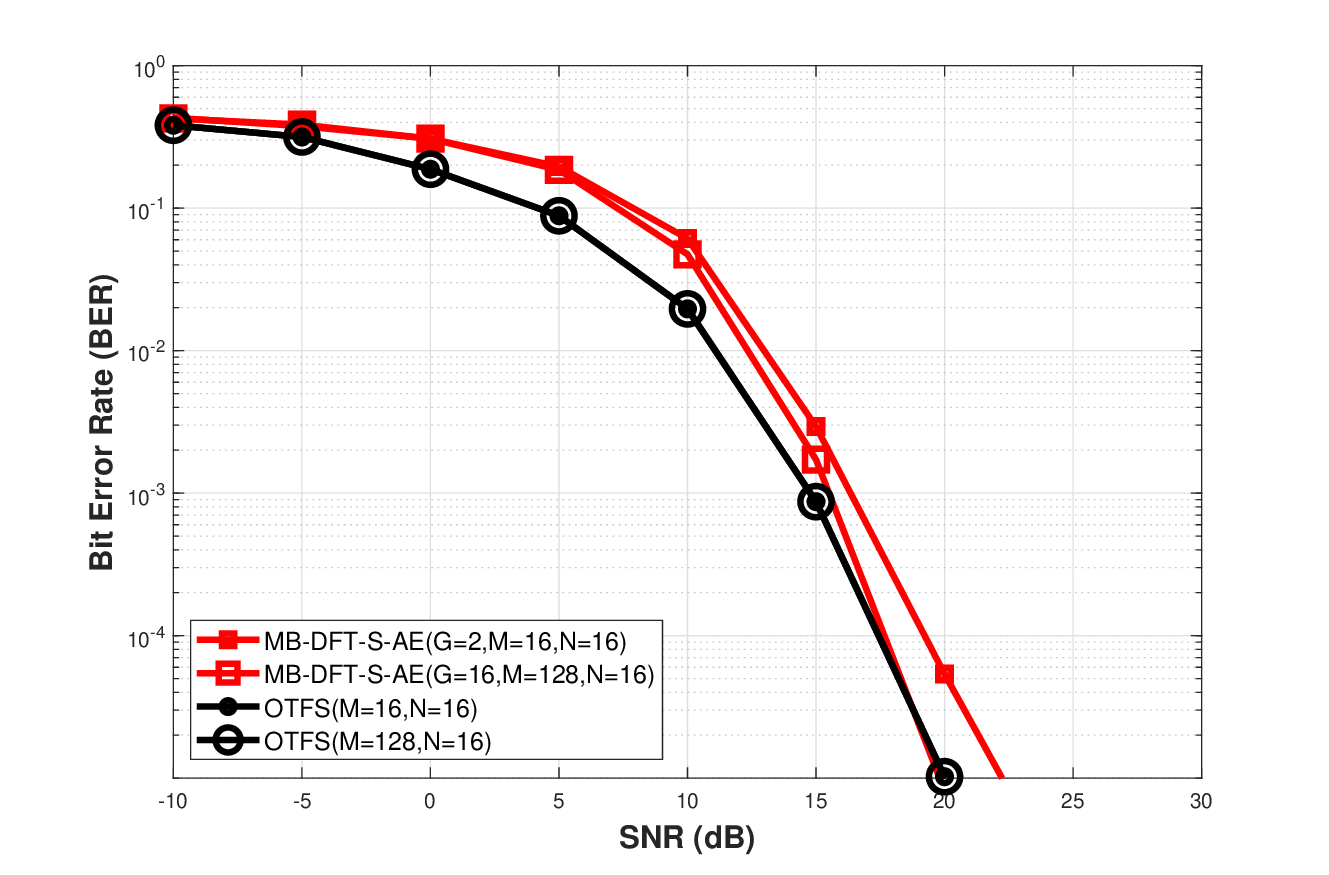}    
\label{fig:MB} 
\end{minipage}}
\subfigure[BER performance of MB-AE-DFT-S-OTFS with different $\eta$. ]{                    
\begin{minipage}{5.6cm}
\includegraphics[width=5.4cm]{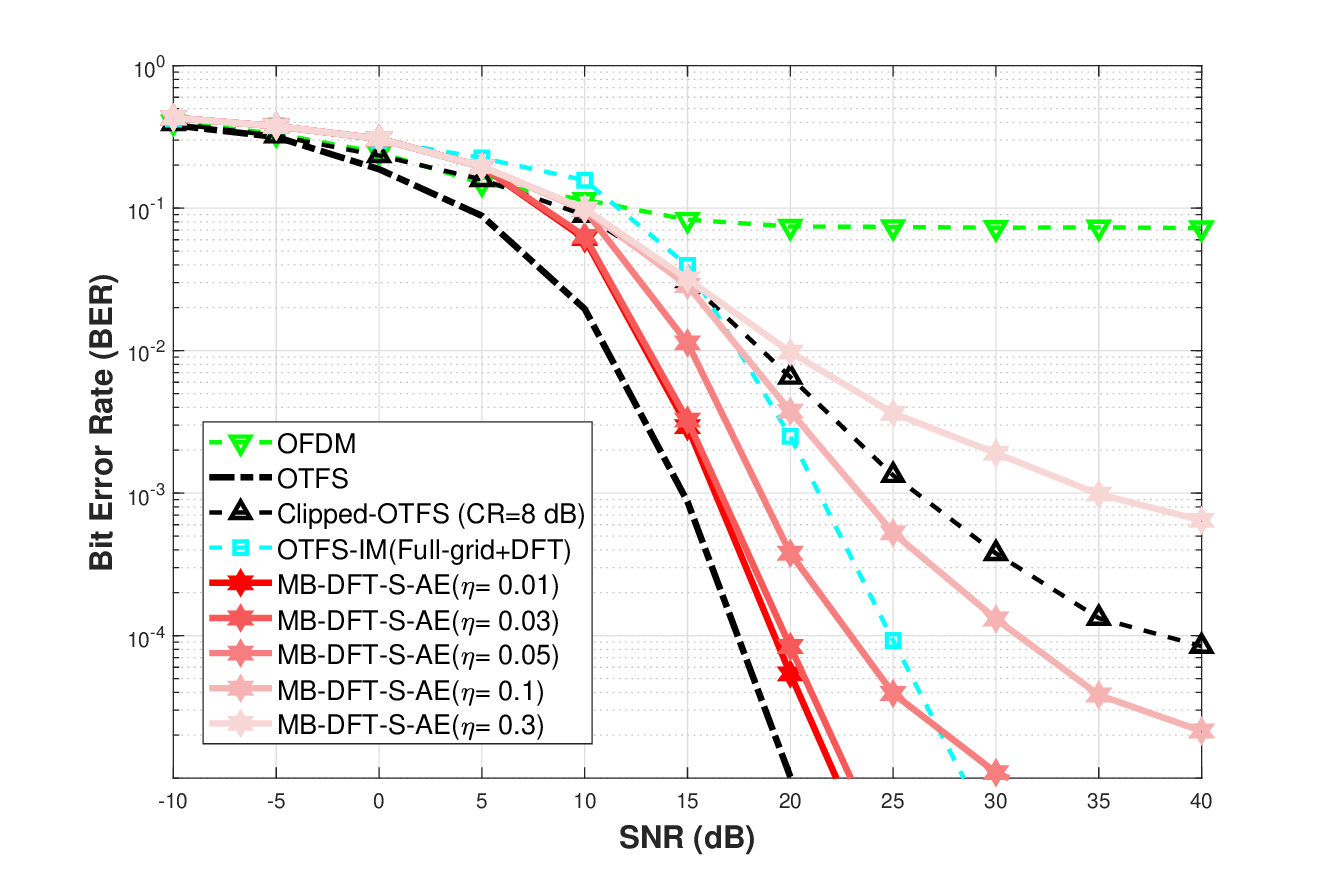}
\label{fig:s41_eta_ber}
\end{minipage}}
\subfigure[BER performance of MB-AE-DFT-S-OTFS with different trained SNR. ]{ 
\begin{minipage}{5.6cm}
\includegraphics[width=5cm]{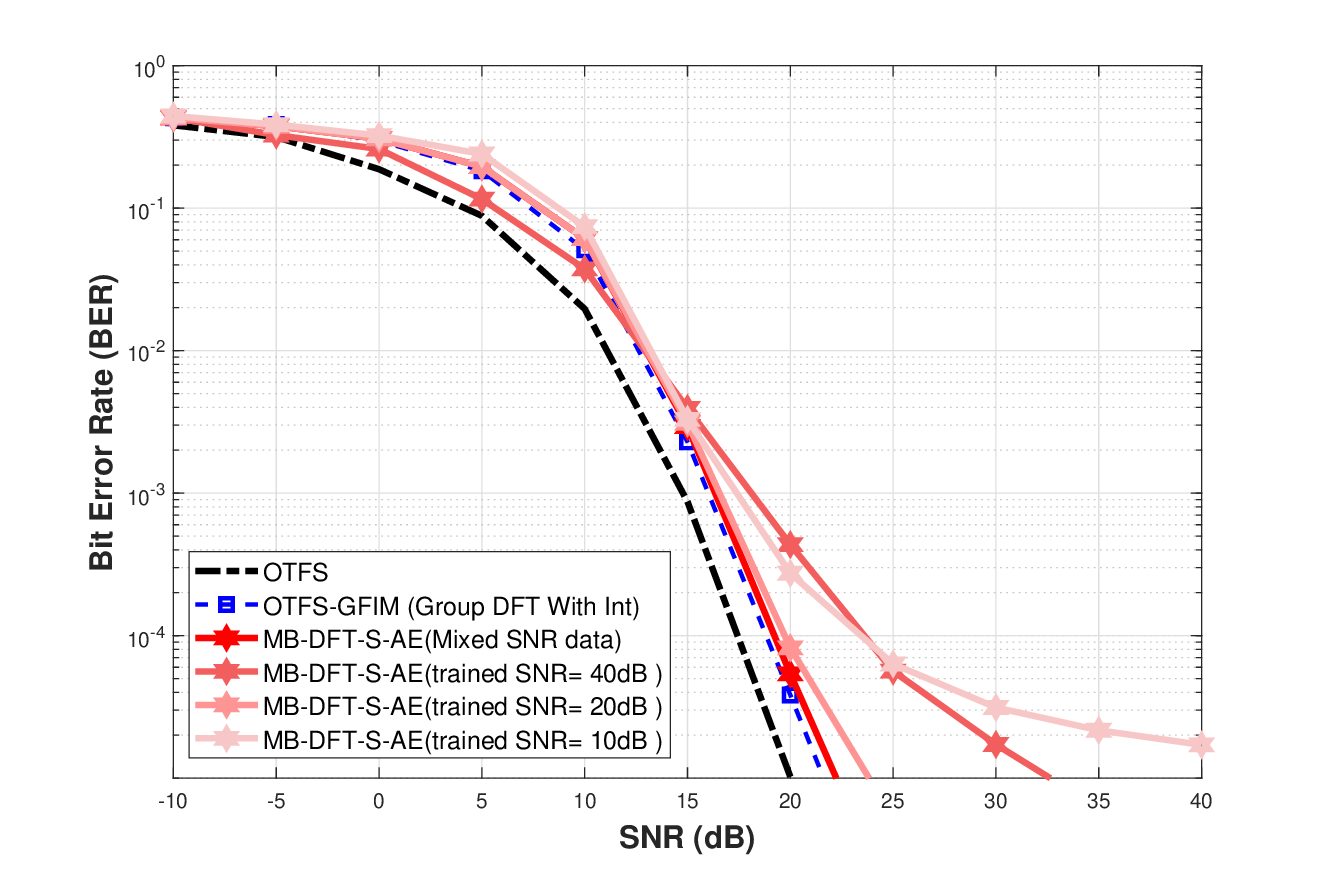}
\label{fig:trainSNR}
\end{minipage}}
\centering                          
\subfigure[CCDF of PAPR of MB-AE-DFT-S-OTFS in comparison to OTFS.]{                    
\begin{minipage}{5.6cm}
\includegraphics[width=5.4cm]{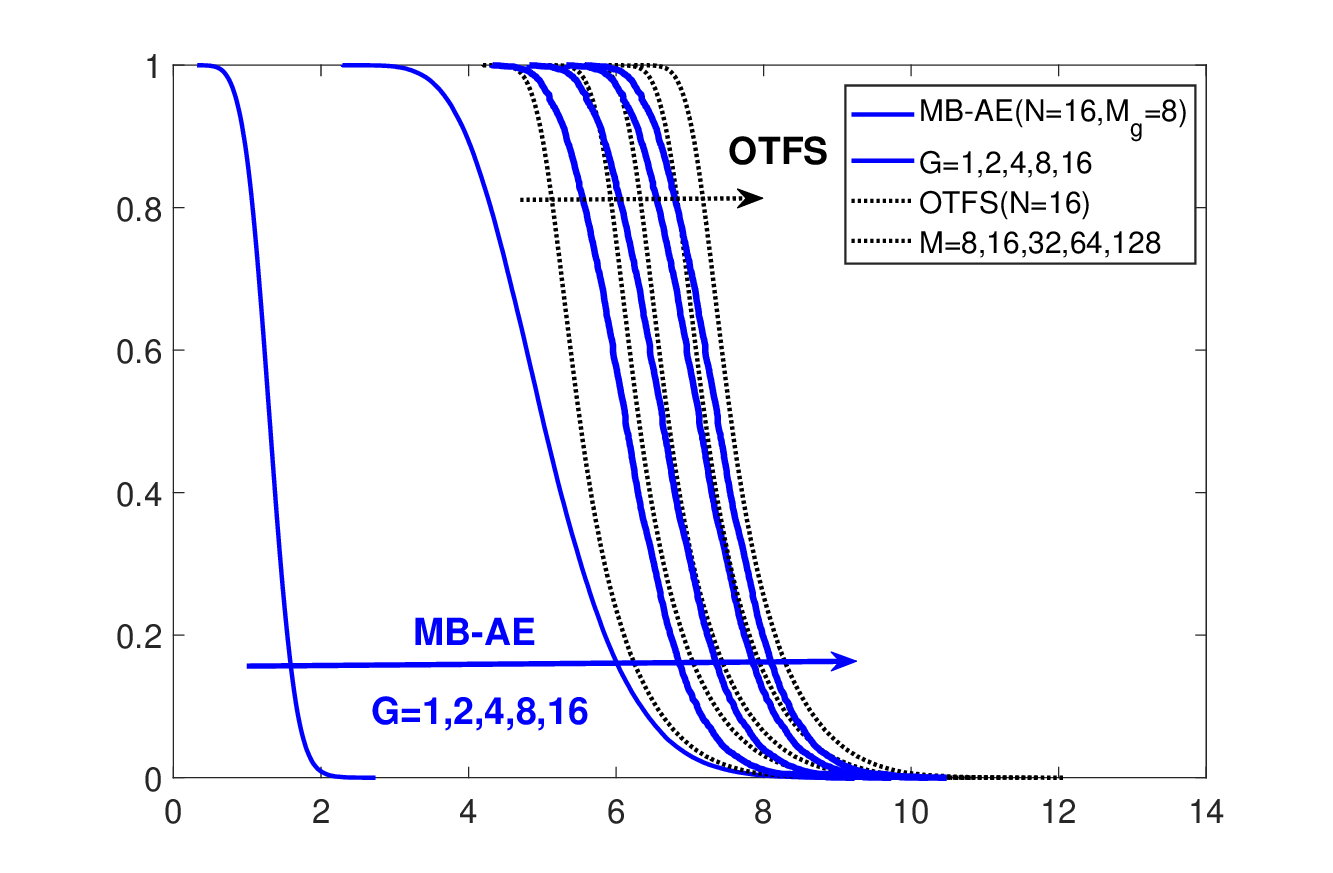}    
\label{fig:MB2} 
\end{minipage}}
\subfigure[CCDF of PAPR of MB-AE-DFT-S-OTFS with different $\eta$.]{                    
\begin{minipage}{5.6cm}
\includegraphics[width=5.4cm]{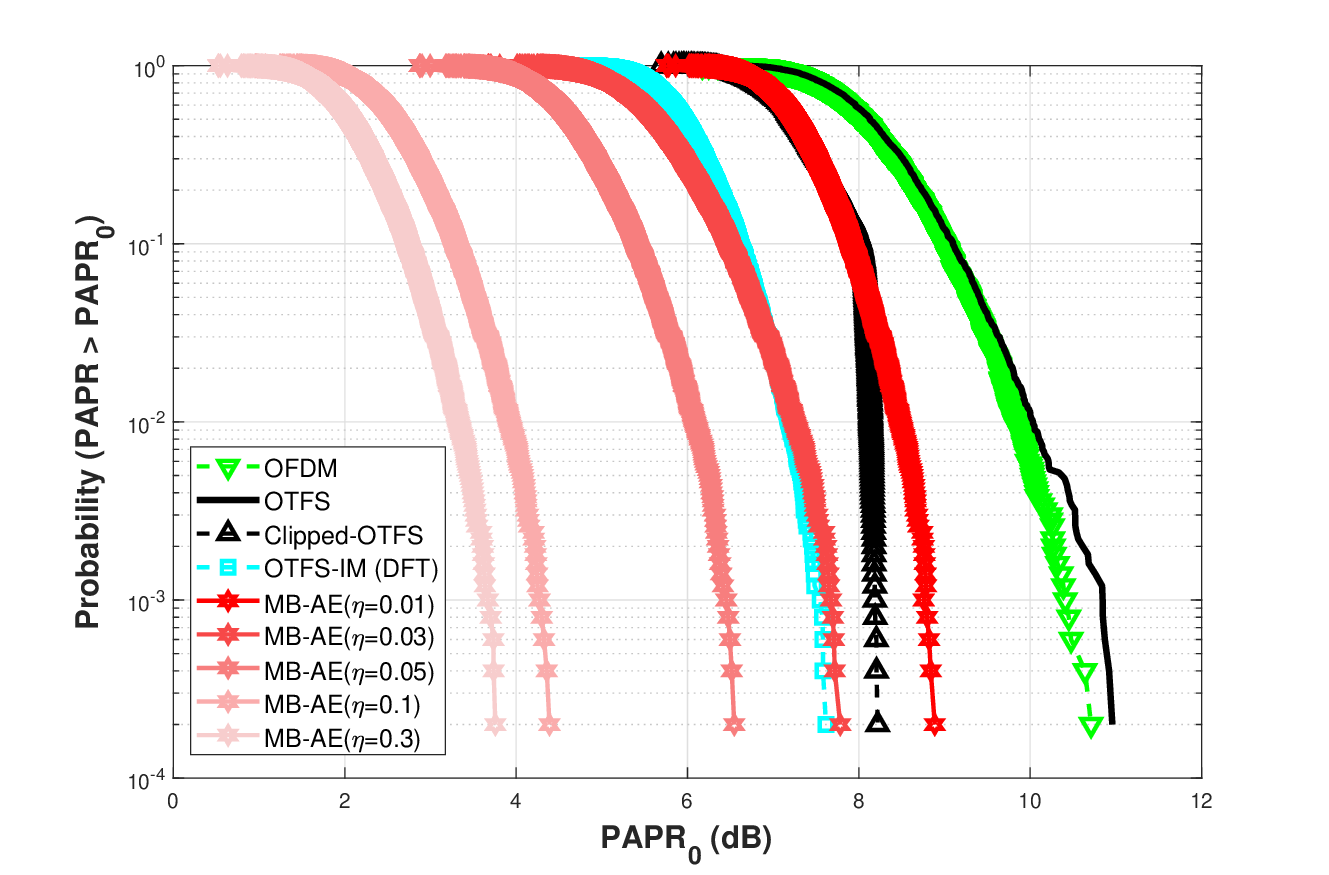}
\label{fig:s42_eta_papr}
\end{minipage}}
\subfigure[BER performance of MB-AE-DFT-S-OTFS with different channel estimation error.]{ 
\begin{minipage}{5.6cm}
\includegraphics[width=5.2cm]{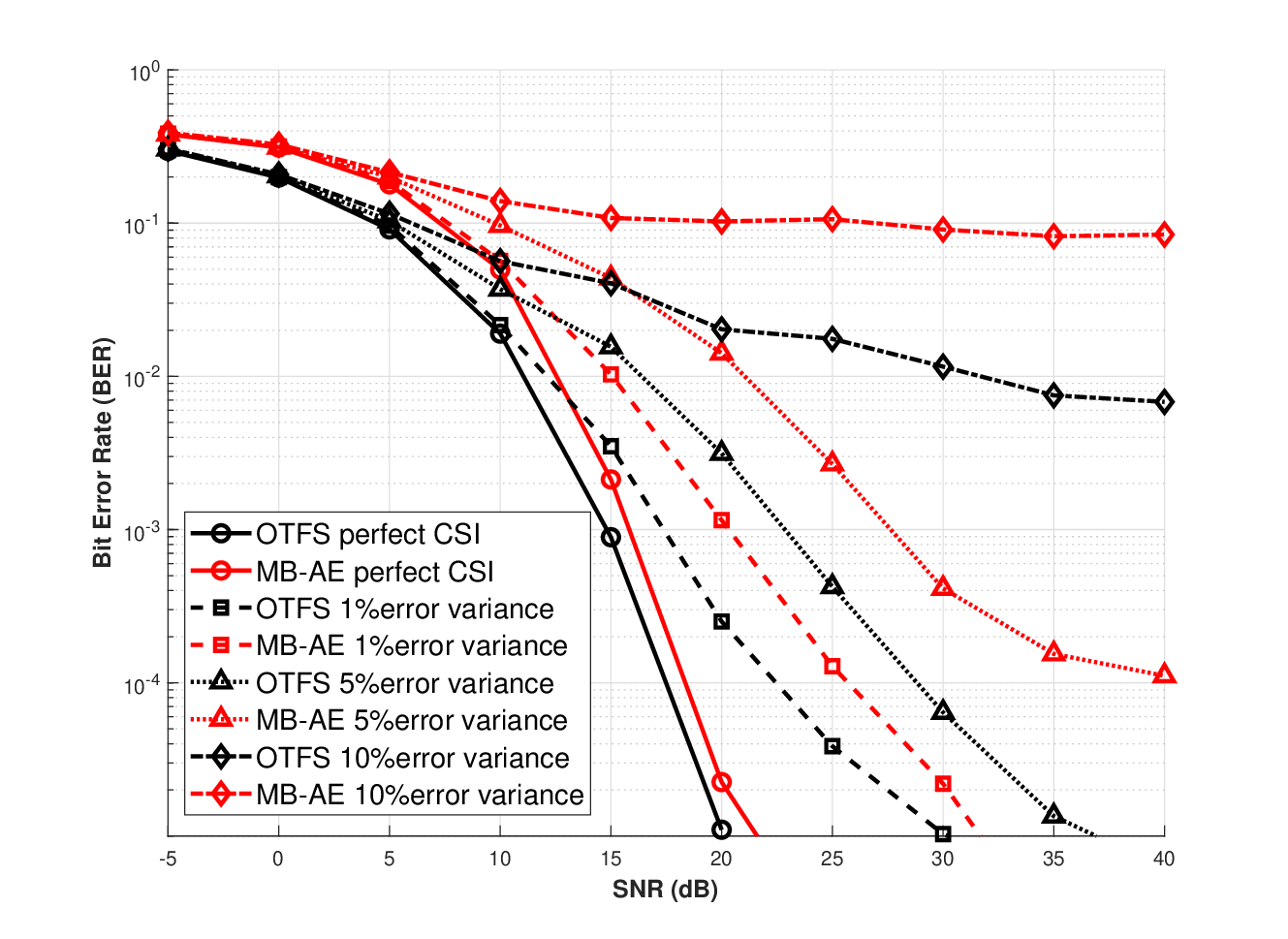}
\label{fig:ce11}
\end{minipage}}
\centering
\caption{Illustration of the parameters and setting configuration of  the proposed MB-AE-based scheme  over a shadowed-Rician NTN channel}  
\vspace{-6mm}
\end{figure*}

Regarding the index multiplexing and detection complexity, traditional full-grid {OTFS-IM} requires a combinatorial search across the entire Delay-Doppler grid. With $K_{z,\text{full}}$ silent subcarriers, the optimal Maximum Likelihood (ML) detection scales as $\mathcal{O}(\binom{MN}{K_{z,\text{full}}} \cdot MN)$. Even for a single silent subcarrier, this results in an $\mathcal{O}((MN)^2)$ polynomial complexity, which rapidly deteriorates into an exponential "curse of dimensionality" as $K_z$ increases. The proposed Multi-Band architecture circumvents this by partitioning the grid into $G$ independent sub-bands of size $N_f$. The combinatorial search is thus strictly confined within localized blocks, drastically reducing the detection complexity to a linearly scalable $\mathcal{O}(G \cdot \binom{N_f}{K_{z,\text{gfim}}} \cdot N_f)$. Furthermore, the online inference complexity of the Autoencoder ({AE}) must be explicitly considered. For a conventional full-grid {AE}, the dense matrix multiplications within the fully connected layers necessitate a computational complexity of $\mathcal{O}((MN)^2)$. In stark contrast, the proposed {MB-AE} decomposes this global non-linear shaping into $G$ parallelized sub-networks. The inference complexity per sub-band is strictly bounded to $\mathcal{O}(N_f^2)$, yielding a total global complexity of $\mathcal{O}(G \cdot N_f^2) = \mathcal{O}(G \cdot (MN/G)^2)$. 

To characterize this scalability and robustness empirically, we compare three configurations: $M=64$ with $4$-QAM (Fig.~\ref{fig:s11_ber}, \ref{fig:s12_papr}), $M=128$ with $4$-QAM (Fig.~\ref{fig:s21_ber}, \ref{fig:s22_papr}), and $M=16$ with $16$-QAM (Fig.~\ref{fig:s31_ber}, \ref{fig:s32_papr}). 

First, comparing $M=64$ and $M=128$ demonstrates the impact of grid scaling. As expected, when $M$ increases, the {PAPR} of conventional {OTFS} (at a CCDF of $10^{-3}$) increase from $10.5$~dB to over $11.0$~dB. By contrast, the {MB-AE} grid partitioning prevents this escalation, maintaining a stable {PAPR} of $6.5 \sim 6.6$~dB across both sizes at an BER of $>4$~dB reduction over standard {OTFS}. Simultaneously, {MB-AE} perfectly tracks the theoretical {OTFS} {BER} baseline (Fig.~\ref{fig:s11_ber}, \ref{fig:s21_ber}), whereas Clipped {OTFS} exhibits severe error floors above an SNR of $12$~dB, and the full-grid {AE} fails entirely.

Next, we test adaptability to higher throughput using $M=16$ with $16$-QAM. While higher-order {IM} typically causes severe error floors in full-grid {IM} or compressed {AE}s, the non-linear shaping of {MB-AE} optimally redistributes signal energy to preserve distance of QAM constellation. Consequently, {MB-AE} achieves a {BER} of $10^{-4}$ at $16$~dB SNR, matching the unconstrained {OTFS} baseline in Fig.~\ref{fig:s31_ber}, while simultaneously reducing  the {PAPR} from $8.8$~dB to $5.5$~dB in Fig.~\ref{fig:s32_papr}. This confirms that {MB-AE} is both scalable and robust under high-order modulations.

In this subsection, we characterize the effect of diverse $\eta$ and the robustness of the {MB-AE} architecture. First, we investigate the joint loss weighting parameter $\eta$. As depicted in Fig.~\ref{fig:s41_eta_ber} and \ref{fig:s42_eta_papr}, setting $\eta=0.3$ prioritizes {PAPR} minimization achieving $5.4$~dB at a CCDF of $10^{-3}$ but tather substantially degrades the {BER}. Conversely, $\eta=0.001$ perfectly preserves the {BER} but yields an insufficient {PAPR} reduction about $7.8$~dB. The best position of $\eta=0.1$ delivers a highly competitive {PAPR} of $6.2$~dB without imposing any noticeable {BER} degradation.

Fig.~\ref{fig:trainSNR} illustrates the system's sensitivity to the training {SNR}. Training at a low {SNR} at about $0$~dB forces an overly conservative constellation mapping, inadvertently causing an error floor near $10^{-4}$ at high-{SNR}s. Conversely, training in a near-noiseless environment at about $20$~dB fails to equip the network with resilience against severe {NTN} fading dips. Consequently, the results demonstrate that adopting a mixed-{SNR} training strategy—where training samples dynamically span the entire operational {SNR} range—provides the most robust feature extraction and superior generalization, successfully avoiding the respective performance bottlenecks of fixed training {SNR}s.

Furthermore, we assess the framework's robustness under imperfect {CSI}, a critical challenge in dynamic {LEO} links. As shown in Fig.~\ref{fig:ce11}, introducing a severe estimation error variance of $\sigma_e^2=0.05$ degrades the {BER} by roughly one order of magnitude at an {SNR} of $15$~dB. However, the {MB-AE} detector prevents catastrophic failure, resulting in a stable waterfall curve. This confirms that the neural network's non-linear feature extraction capability inherently mitigates partial channel uncertainty.

Despite this inherent robustness, {AE}-based {HD} frameworks limit bottleneck the achievable channel coding gain under extreme channel impairments. To fully exploit modern {NTN} standards, a paradigm shift toward {SD} architectures is required, where the {AE} directly outputs probabilistic {LLR}s. This extension is comprehensively detailed in the following section.

\begin{figure}[!htb]
    \centering
    \vspace{-2mm}
    \includegraphics[width=0.36\textwidth]{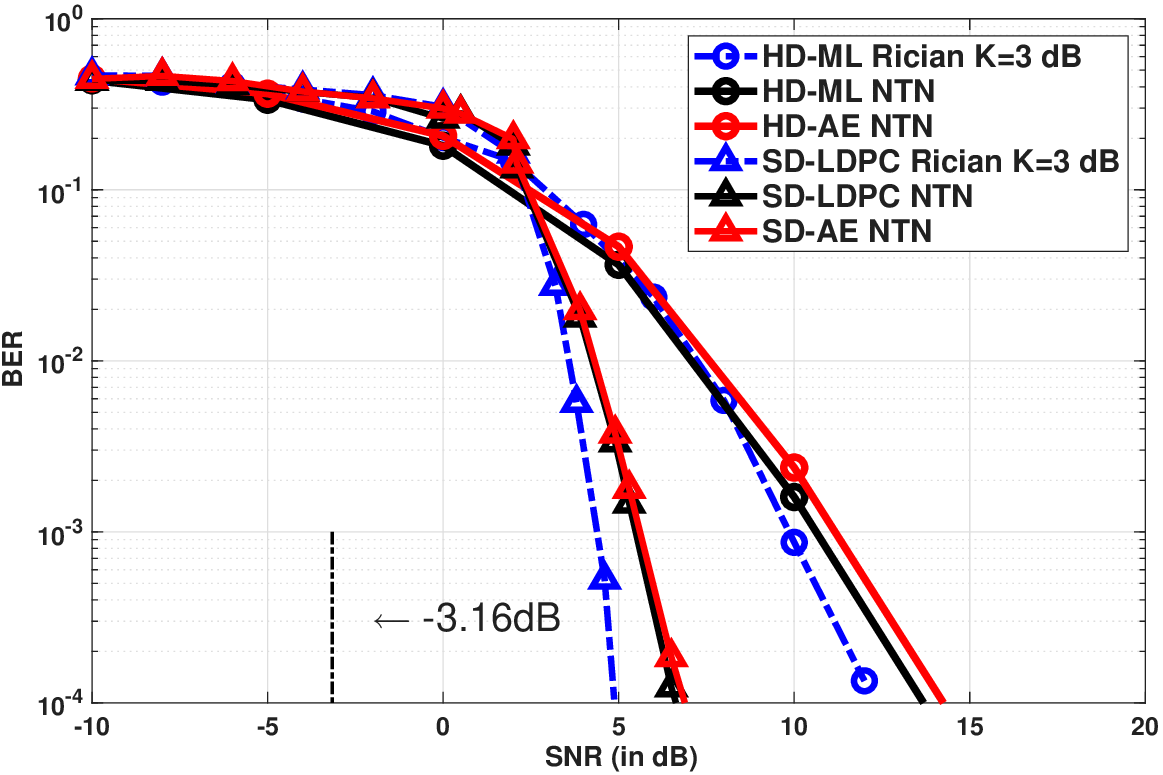}
    \caption{{BER} performance of the proposed AE-based scheme for both HD and SD detection under the Rician based {NTN} channel and the proposed shadowed {NTN} channel with $m'=2$ at an effective throughput of 0.6667 bits/s/Hz.}
    \label{fig:w1}
    \vspace{-4mm}
\end{figure}
\begin{figure}[!t]
    \centering
    \vspace{-4mm}
    \includegraphics[width=0.36\textwidth]{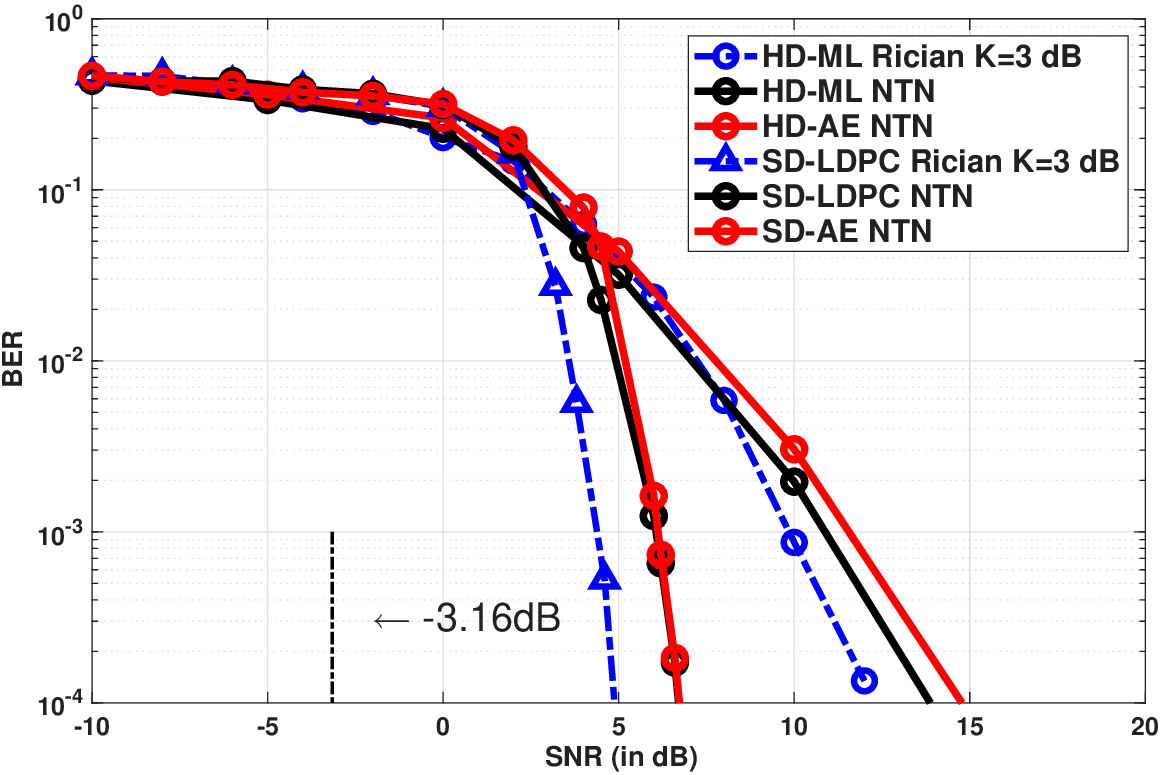}
    \caption{{BER} performance of the proposed AE-based scheme for both HD and SD detection under the Rician based {NTN} channel and the proposed shadowed {NTN} channel with $m'=4$ at an effective throughput of 0.6667 bits/s/Hz.}
    \label{fig:w2}
    \vspace{-2mm}
\end{figure}
\begin{figure}[!t]
    \centering
    \vspace{-4mm}
    \includegraphics[width=0.36\textwidth]{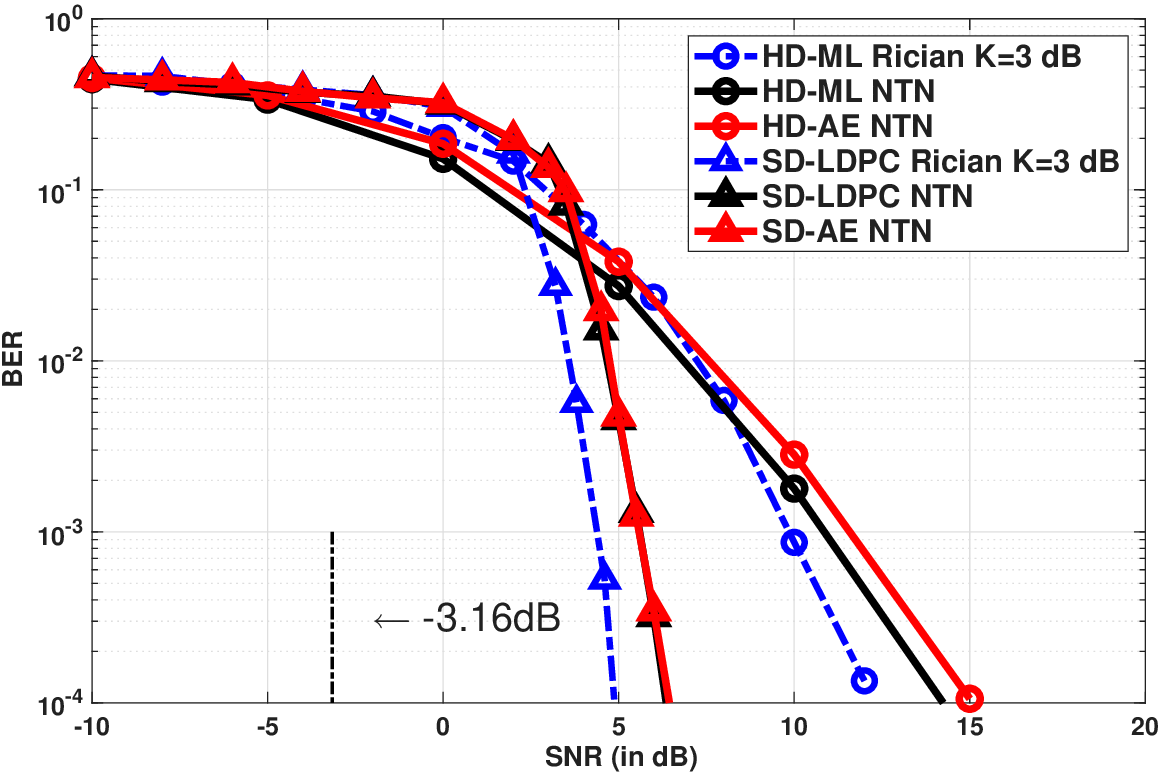}
    \caption{{BER} performance of the proposed AE-based scheme for both HD and SD detection under the Rician based {NTN} channel and the proposed shadowed {NTN} channel with $m'=16$ at an effective throughput of 0.6667 bits/s/Hz}
    \vspace{-3mm}
    \label{fig:w3}

\end{figure}

Fig. \ref{fig:w1}- \ref{fig:w3} compares the {BER} performance of the proposed system using HD Maximum Likelihood  and SD detection along with {LDPC} channel coding scheme under a  Rician {NTN} channel associated with $K=3$ and for transmission over a proposed shadowed channel based on  different Nakagami-$m$ values, corresponding to different satellite visibility/elevation conditions. 
Under Rician fading, the SD-LDPC detector exhibits the best {BER} performance, providing a substantial coding gain relative to its HD counterparts as shown in Fig. \ref{fig:w1}.\par
Furthermore, as observed in Figs.~\ref{fig:w1}-\ref{fig:w3}, the {HD-AE} and {SD-AE} curves remain close to conventional approaches. The  {NTN} channel model employed introduces characteristic shadowing effects, which are parameterized by the Nakagami shape factor $m'$. A smaller $m'$ corresponds to heavy blockage and lower satellite elevation angles as characterized in Fig.~\ref{fig:w1}, whereas a higher $m'$ represents mild shadowing conditions onFig.~\ref{fig:w3}. Despite operating with fixed network weights, the {SD-AE} scheme demonstrates clear performance improvements over both {HD-AE} and conventional HD-Maximum Likelihood approaches across all $m'$ values employed. This confirms that the non-linear features learned exhibit exceptional generalization capabilities, allowing the {AE} to retain strong robustness against unpredictable, varying shadowed environments when combined with soft decoding.

\vspace{-4mm}
\section{CONCLUSIONS}

A DFT-spread autoencoder-based PAPR reduction scheme was proposed for {OTFS-IM} modulation under a practical {NTN} channel. The proposed framework integrates DFT spreading with a deep AE architecture by exploiting a learning-based autoencoder. The encoder maps the input {OTFS-IM} signal into a low-PAPR representation, while the decoder reconstructs the original signal with minimal {BER} degradation. To facilitate joint optimization, we designed a novel loss function that incorporates both the reconstruction accuracy and PAPR minimization, while relying on a tunable weighting parameter to balance the {BER} \textit{vs.} power efficiency trade-off. A two-stage training strategy is employed for improving convergence, involving pretraining without noise and fine-tuning under realistic SNR conditions. Our simulation results demonstrated that the proposed DFT-spread AE-based {OTFS-IM} scheme significantly outperforms conventional PAPR reduction techniques in terms of both PAPR suppression and {BER} robustness, which could be further enhanced by harnessing more advanced detectors. These results highlight the effectiveness of combining IM, DFT spreading, and deep learning for next-generation high-mobility {NTN} scenarios.The AE learns the most appropriate joint configurations of the system components, beneficially annealing them into a potent amalgamated transceiver.

% or


%\appendices

%\section{}

%\section*{Acknowledgment}


\ifCLASSOPTIONcaptionsoff
  \newpage
\fi



\vspace{-3mm}
\bibliographystyle{IEEEtran}
\bibliography{ref}


%\end{IEEEbiography}


%\newpage



%\vfill


\end{document}